\documentclass[preprint2]{aastex}
\usepackage{longtable}
\usepackage{natbib}
\usepackage{amsmath}
\usepackage[colorlinks,linkcolor=blue,anchorcolor=blue,citecolor=blue]{hyperref}
\usepackage{ulem}

\shorttitle{
Calibration of LAMOST $\log{g}$
}
\shortauthors{Wang et al.}

\begin{document}

\title{Calibration of LAMOST Stellar Surface Gravities Using the {\em Kepler}
    Asteroseismic Data
}

\author{
    Liang Wang\altaffilmark{1,2},
    Wei Wang\altaffilmark{1,3},
    Yue Wu\altaffilmark{1},
    Gang Zhao\altaffilmark{1},
    Yinbi Li\altaffilmark{1},
    Ali Luo\altaffilmark{1},
    Chao Liu\altaffilmark{1},
    Yong Zhang\altaffilmark{4}, 
    Yonghui Hou\altaffilmark{4}, 
    Yuefei Wang\altaffilmark{4} 
    \and
    Zihuang Cao\altaffilmark{1} 
}

\affil{
}
\email{[lwang;wangwei;gzhao]@nao.cas.cn}

\altaffiltext{1}{
    Key Laboratory of Optical Astronomy,
    National Astronomical Observatories, Chinese Academy of Sciences,
    A20, Datun Road, Chaoyang District, Beijing 100012, China
}
\altaffiltext{2}{
    Max-Planck-Institut f\"ur Extraterrestrische Physik,
    Giessenbachstrasse,
    85748 Garching, Germany
}
\altaffiltext{3}{
    Chinese Academy of Sciences South America Center for Astronomy,
    Camino El Observatorio 1515, Las Condes, Santiago, Chile
}
\altaffiltext{4}{
    Nanjing Institute of Astronomical Optics \& Technology,
    National Astronomical Observatories, Chinese Academy of Sciences,
    Nanjing 210042, China
}

\date{Received ; accepted}

\begin{abstract}
Asteroseismology is a powerful tool to precisely determine the evolutionary
    status and fundamental properties of stars.
With the unprecedented precision and nearly continuous photometric data acquired
    by the NASA {\em Kepler} mission, parameters of more than $10^4$ stars have
    been determined nearly consistently.
However, most studies still use photometric effective temperatures
    ($T_{\rm eff}$) and metallicities ([Fe/H]) as inputs, which are not
    sufficiently accurate as suggested by previous studies.
We adopted the spectroscopic $T_{\rm eff}$ and [Fe/H] values based on the LAMOST
    low-resolution spectra ($R\simeq1,800$), and combined them with the global
    oscillation parameters to derive the physical parameters of a large sample
    of stars.
Clear trends were found between $\Delta\log{g}({\rm LAMOST}-{\rm seismic})$ and
    spectroscopic $T_{\rm eff}$ as well as $\log{g}$, which may result in an
    overestimation of up to 0.5\,dex for the $\log{g}$ of giants in the LAMOST
    catalog.
We established empirical calibration relations for the $\log{g}$ values of
    dwarfs and giants.
These results can be used for determining the precise distances to these stars
    based on their spectroscopic parameters.

\end{abstract}


\section{INTRODUCTION}
Wide-field, multi-object spectroscopic surveys such as the Sloan Extension for
    Galactic Understanding and Exploration \citep[SEGUE;][]{Yanny2009}, RAdial
    Velocity Experiment \citep[RAVE;][]{Steinmetz2006}, and Large sky Area
    Multi-Object fiber Spectroscopic Telescope \citep[LAMOST;][]{Cui2012}, have
    proved to be efficient for exploring the Milky Way galaxy.
Determining the fundamental parameters and chemical characteristics of a large
    sample of stars is particularly important and essential for better
    understanding the formation and structure of galaxies.
Stellar surface gravity, $\log{g}$, is one of the most crucial parameters in
    stellar physics as it is closely related to the stellar luminosity and,
    hence, to the position of a star on the Hertzsprung-Russell Diagram (HRD).
In addition, if the stellar mass is known, one can obtain the stellar radius and
    reddening-independent distance with precision superior to that of
    photometric calibrations \citep[e.g.][]{Breddels2010,Xue2014}.
On the other hand, precise determination of magnesium and calcium abundances
    from Mg\,Ib and infrared Ca\,II triplets in low-resolution spectra rely
    heavily on the accurate determination of $\log{g}$ \citep[e.g.][]{
    Deeming1960,Chmielewski2000}.

In the high-resolution ($R>40,000$) spectroscopy, several approaches are often
    used for determining the $\log{g}$ values of cool stars.
The first approach utilizes the ionization balance of neutral and singly ionized
    atoms of the same element, such as Fe I/II \citep[e.g.][]{Fuhrmann1998,
    daSilva2006, Boesgaard2011}.
A typical $\log{g}$ error in this approach is 0.1-0.2\,dex, which is limited by
    the facts that
    (1) the number of unblended, weak, singly ionized iron lines in stellar
    spectra is too small;
    (2) the equilibria of Fe I and Fe II are strongly affected by
    $T_{\rm eff}$; and
    (3) the non-local thermodynamic equilibrium (NLTE) effect affects the
    abundance of neutral iron lines by up to 0.1\,dex \citep[e.g.][]{
    Mashonkina2011, Lind2012}.
A different method for determining $\log{g}$ uses the basic relation
    $\log{g}=\log{M}+4\log{T_{\rm eff}}+0.4M_{\rm bol}$ \citep[e.g.][]{Chen2000,
    Reddy2003, Wang2011}, where the accurate absolute bolometric magnitude
    $M_{\rm bol}$ relies on the data of precise trigonometric parallaxes (e.g.,
    acquired by the {\sc Hipparcos} mission).
A relative parallax uncertainty of 20\% yields an error of 0.17\,dex in
    $\log{g}$.
In the {\sc Hipparcos catalogue}, $\sim$60\% of the stars with distances above
    100\,pc are characterized by a relative parallax uncertainty above 20\%
    \citep{vanLeeuwen2007}.

Determination of stellar atmospheric parameters ($T_{\rm eff}$, $\log{g}$, and
    [Fe/H]) from low- to medium-resolution spectra are mostly based on the
    spectral synthesis technique, with a library covering a wide range of
    $T_{\rm eff}$, $\log{g}$, and [M/H] values \citep[e.g.][]{Zwitter2004,
    Prungniel2001,Sanchez2006, Cenarro2007}.
The precision associated with $\log{g}$ is generally lower than that obtained in
    high-resolution spectroscopy.
For example, the error on $\log{g}$ determined from the SEGUE Stellar Parameter
    Pipeline \citep[SSPP;][]{Lee2008} is $\sim$0.23\,dex, while it is 0.5\,dex
    for RAVE \citep{Zwitter2008}.
The precision of $\log{g}$ for the ongoing LAMOST survey is $\sim0.2$\,dex for
    both the LAMOST stellar parameter pipeline \citep[LASP,][]{Wu2014,Luo2015}
    and LSP-3 \citep{Xiang2015,Ren2016}.
\cite{Carlin2015} developed a Bayesian model to derive stellar distances from
    calibrated stellar spectra, and applied it to the LAMOST data.
They found that the precision with which distances could be determined was
    limited to 40\% owing to large uncertainties associated with $\log{g}$.
Reducing the $\log{g}$ uncertainty by 0.1\,dex would increase the distance
    accuracy by $\sim$12\% \citep[e.g.][]{Liu2015}.

Launched in 2009 March, the NASA {\em Kepler} space telescope
    \citep{Borucki2010} uses a wide-field, 95-cm-aperture telescope to search
    for transiting Earth-sized planets in a sample of $\sim$170,000 stars.
The data collected during the first four years of the operation of this
    telescope not only revolutionized the extra-solar planet hunting campaign
    but also significantly contributed to other fields, such as
    asteroseismology.
With the unprecedented photometric precision, researchers are, for the first
    time, able to precisely determine the $M$, $R$, $\log{g}$, and $\rho$ values
    for $\sim10^4$ stars by consistently using the asteroseismology method
    \citep[e.g.][]{Kallinger2010, Hekker2011, Stello2013, Chaplin2014,
    Huber2014}.
These stars reveal solar-like oscillations in their power spectra, and their
    spectral types range from early F to late K \citep{Chaplin2013}, including
    both giants and dwarfs.
The uncertainty associated with asteroseismic $\log{g}$ is typically less than
    0.02\,dex, which is one order of magnitude lower than the spectroscopically
    determined one \citep{Hekker2013}.
\cite{Gai2011} showed that such asteroseismically determined $\log{g}$ values
    are almost independent of the stellar evolution model grid and contain
    nearly no systematic errors.

Derivation of stellar physical parameters ($M$, $R$, $L$) using
    asteroseismology scaling relations relies on $T_{\rm eff}$ and [Fe/H] from
    ``external" sources as inputs.
The majority of asteroseismically interesting stars in the {\em Kepler} field
    \citep[e.g.][]{Chaplin2014} are analyzed by adopting photometric or
    Infra-Red Flux Method (IRFM) calibrated $T_{\rm eff}$, together with Kepler
    Input Catalog \citep[KIC;][]{Brown2011} metallicities, in which systematic
    errors or large scatter have already been found \citep[e.g.][]{Dong2014}.
For these stars, $T_{\rm eff}$ and [Fe/H] based on high-resolution spectroscopy
    remain a challenge because most of these stars are too faint for
    modest-sized telescopes.
Recently, a significant amount of data on low-resolution ($R\sim1,800$) spectra
    in the {\em Kepler} field have been released by the LAMOST survey
    \citep{Luo2015}, and a set of consistent, spectroscopic $T_{\rm eff}$ and
    [Fe/H] values has been reliably determined.
Therefore, it is worthwhile to redetermine the physical parameters of these
    stars by replacing the photometric or KIC inputs by this new set of
    atmospheric parameters.

\section{LAMOST SPECTROSCOPIC DATA}

\subsection{LAMOST Observations in the {\em Kepler} Field}
LAMOST, also known as the ``Guoshoujing Telescope,'' is a reflecting Schmidt
    telescope with an effective aperture of $\sim$4\,m and a field of view (FOV)
    of 20\,deg$^2$.
Four thousand fiber units in its focal plane and 16 multi-object spectrographs
    make it highly efficient for spectroscopic surveys.
During the first three years of operation, from 2011 October, to 2014 June,
    LAMOST has collected over 4.1 million spectra with resolving power
    ($R=\lambda/\Delta\lambda$) of 1,800, and public access to these spectra has
    been granted in the second data release
    (DR2)\footnote{\url{http://dr2.lamost.org/}}.
We cross-matched the DR2 and DR3 Quarter 1 (DR3Q1) catalogs with the KIC, and
    found 87,834 spectra of 70,703 common objects within 36 exposures in the
    LAMOST-{\em Kepler} project \citep{DeCat2015}.
Atmospheric parameters for 48,486 stars out of these objects have been
    determined by LASP \citep{Wu2014, Wu2011}.
The median uncertainties of $T_{\rm eff}$, $\log{g}$, and [Fe/H] were 128\,K,
    0.47\,dex, and 0.15\,dex for spectra with signal-to-noise ratios (SNRs) of
    $\sim$50 at 477\,nm, and 101\,K, 0.44\,dex, and 0.12\,dex for spectra with
    SNR of $\sim$100.

\subsection{LAMOST vs. High-Resolution Spectroscopy}

\begin{figure*}[htbp]
    \centering
    \includegraphics[width=15cm]{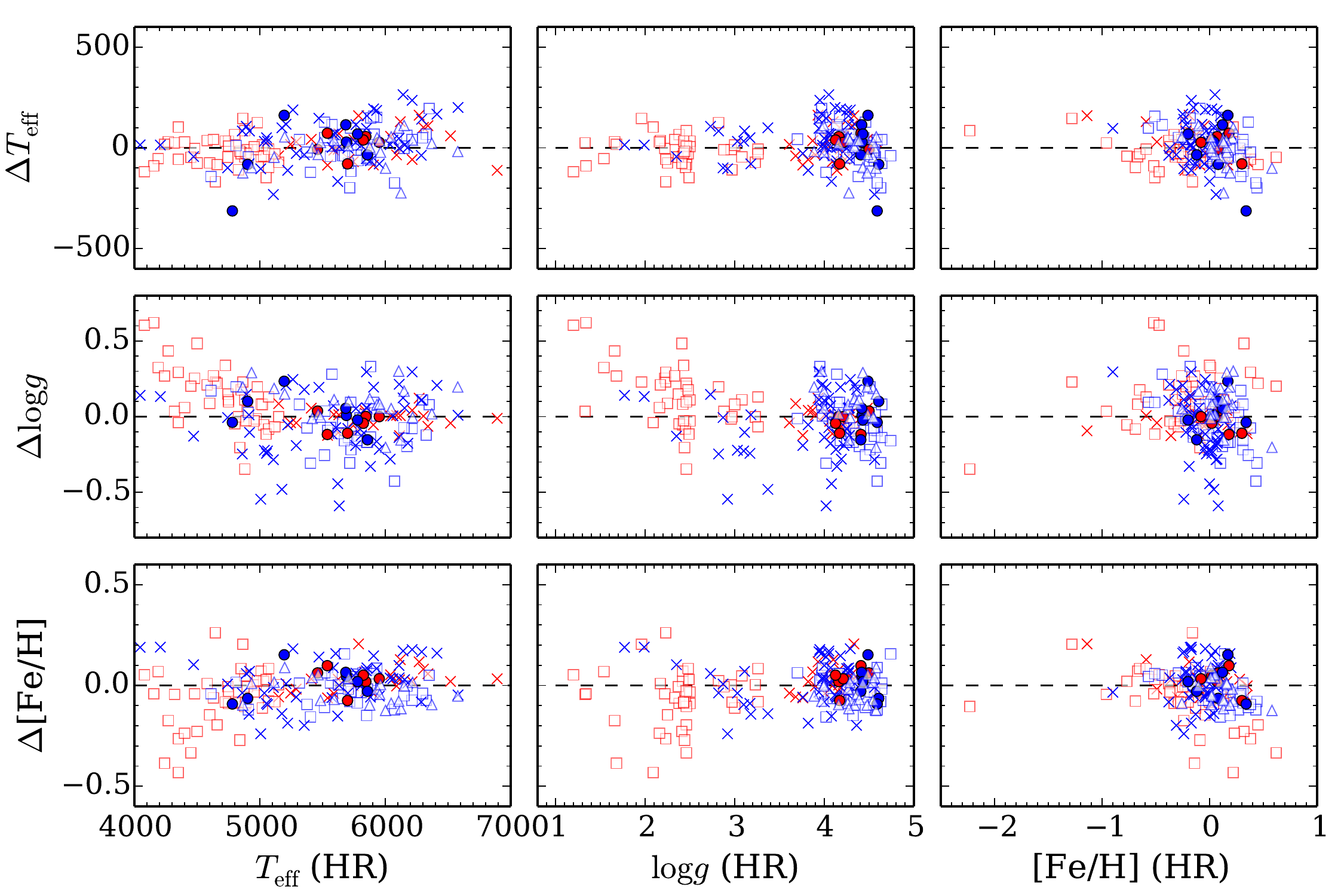}
    \caption{
        Differences between stellar parameters obtained from the LAMOST pipeline
        and those obtained using the HR spectroscopy, as functions of
        $T_{\rm eff}$, $\log{g}$ and [Fe/H], respectively.
        Red circles, red squares, and red crosses represent the LAMOST stars in
        common with M2014 (AST sub-sample), Th2012, and Br2012, respectively.
        All of the above adopted asteroseismic $\log{g}$.
        Blue circles, blue squares, blue crosses and blue triangles are those in
        common with M2014 (SME sub-sample), Bu2012, MZ2013, and Hi2014, all of
        which were obtained using spectral synthesis or excitation/ionization
        equilibrium method.
    }
    \label{comp_lamost_hrs}
\end{figure*}

The asteroseismic scaling relations (see Section~\ref{method}) require
    $T_{\rm eff}$ as an input parameter.
Thus, it is necessary to compare the LAMOST results with those of
    high-resolution spectroscopy (HRS).
However, such a comparison for a large sample of stars is not always feasible,
    because most targets of the LAMOST observing plan are not sufficiently
    bright, and thus, are lacking of HR studies.
Fortunately, the wealth of planet candidate hosts and other stars with
    noticeable values from the {\em Kepler} mission has generated significant
    interest in ground-based follow-up observations, and many of these are
    performed using HR spectrographs on large telescopes, such as the
    10-m-aperture Keck I telescope and the Subaru telescope.
As a result, accurate stellar parameters for hundreds of FGK stars in the
    {\em Kepler} field have been determined using various techniques, providing
    a good opportunity to test the LAMOST low-resolution spectra parameters.

\citet[][hereafter, Br2012]{Bruntt2012} and
\citet[][hereafter, Th2012]{Thygesen2012} observed 93 solar-like and 82 red
    giant stars using high-resolution spectrographs.
They determined the $\log{g}$ values for these stars from global oscillation
    parameters, while the other atmospheric parameters $T_{\rm eff}$, [Fe/H],
    and $\xi$ (micro-turbulent velocity) were determined using the spectroscopic
    method.
\citet[][hereafter, MZ2013]{Molenda2013} also analyzed 169 {\em Kepler} targets
    using spectral synthesis based on high resolution spectra collected by
    different ground-based telescopes.
Moreover, \citet[][hereafter, Bu2012]{Buchhave2012} studied the HR spectra of
    152 planet-host stars using Stellar Parameter Classification (SPC), which is
    also a realization of spectral synthesis with a grid of template spectra.
To validate and characterize the planetary properties, \citet[][hereafter,
    M2014]{Marcy2014} published stellar parameters of 22 {\em Kepler} Objects of
    Interests (KOIs) using the reconnaissance spectra obtained using the HIRES
    spectrometer \citep{Vogt1994}.
\citet[][hereafter, Hi2014]{Hirano2012, Hirano2014} also derived stellar
    parameters for 40 KOIs using the excitation/ionization equilibrium of Fe I
    and Fe II lines.
We divided the above samples into two groups -- depending on how the values of
    $\log{g}$ were derived -- using either the asteroseismology method or purely
    by using spectroscopic techniques.

\begin{figure}[htbp]
    \centering
    \includegraphics[width=8.5cm]{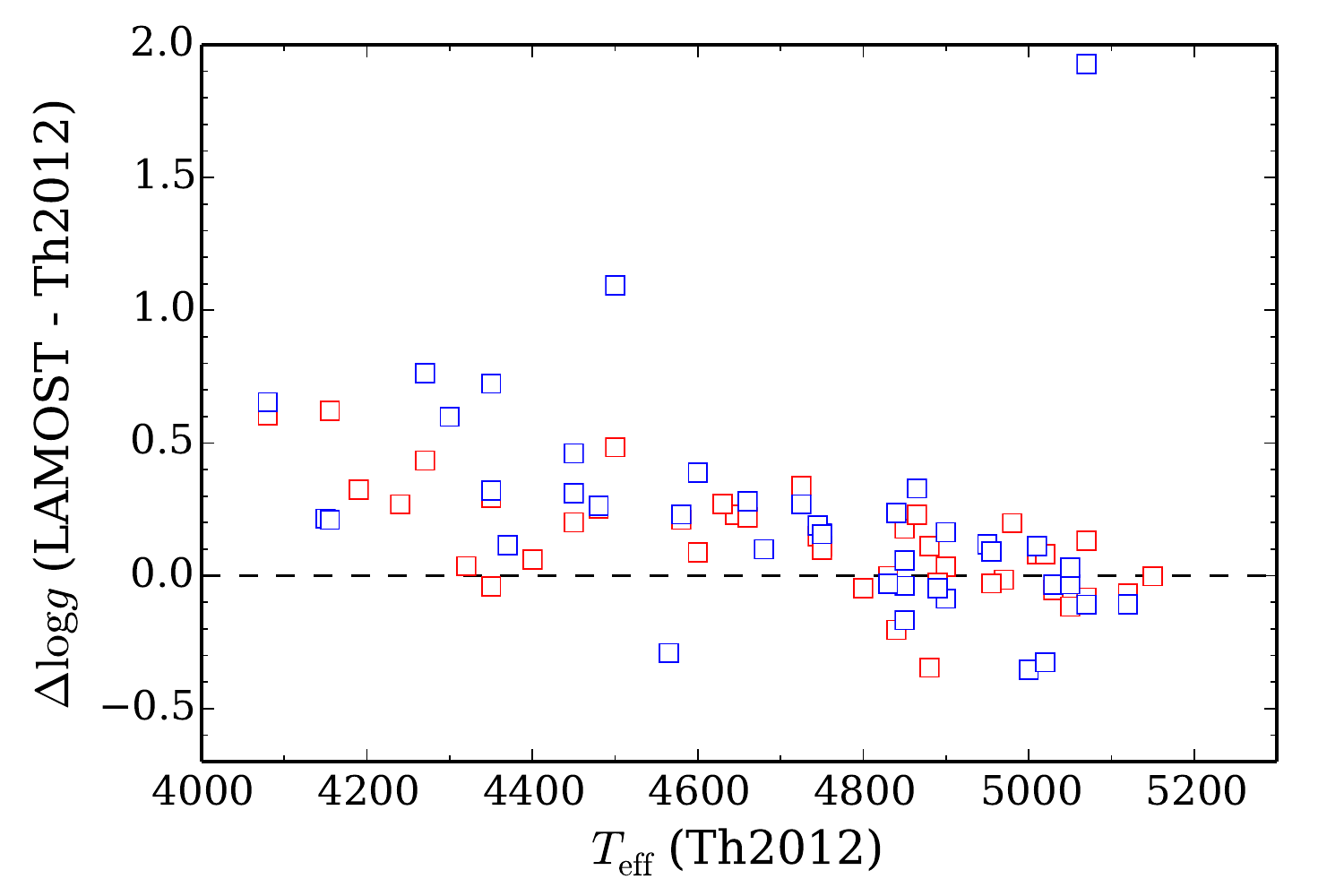}
    \caption{
        Differences between $\log{g}$ from the LAMOST pipeline and from Th2012,
        as functions of $T_{\rm eff}$.
        Similar to Figure~\ref{comp_lamost_hrs}, the red and blue squares
        represent the parameters obtained by performing asteroseismic and
        spectroscopic analysis in Th2012, respectively.
    }
    \label{comp_thygesen}
\end{figure}

Using the LAMOST AFGK-type star parameters catalog, we found 26, 41, 49, 39, 13,
    and 21 common stars with Br2012, Th2012, MZ2013, Bu2012, M2014, and Hi2014,
    respectively.
In Figure~\ref{comp_lamost_hrs}, we compare the stellar parameters extracted
    from literature and the LAMOST catalog.
The mean differences between the LAMOST and high-resolution spectroscopy
    parameters were
    $\left<\Delta T_{\rm eff}\right> =    -1 \pm   71$\,K,
    $\left<\Delta\log{g}     \right> =  0.06 \pm 0.17$\,dex, and
    $\left<\Delta{\rm [Fe/H]}\right> = -0.03 \pm 0.12$\,dex
    for 73 stars in the asteroseismic group (red points in
    Figure~\ref{comp_lamost_hrs}), and
    $\left<\Delta T_{\rm eff}\right> =    19 \pm  100$\,K,
    $\left<\Delta\log{g}     \right> =  0.02 \pm 0.19$\,dex, and
    $\left<\Delta{\rm [Fe/H]}\right> =  0.00 \pm 0.09$\,dex
    for 116 stars in the spectroscopic group (blue points in
    Figure~\ref{comp_lamost_hrs}).
For all the common stars, the mean differences were
    $\left<\Delta T_{\rm eff}\right> =    11 \pm 90$\,K,
    $\left<\Delta\log{g}     \right> =  0.01 \pm 0.18$\,dex, and
    $\left<\Delta{\rm [Fe/H]}\right> = -0.01 \pm 0.10$\,dex.
The temperature values obtained by using the LAMOST catalog were in good
    agreement with those obtained from high-resolution spectra (with
    $\langle\Delta T_{\rm eff}\rangle=-14\pm86$\,K) for stars with
    $T_{\rm eff}<5,500$\,K; however, the difference was slightly higher
    ($31\pm89$ K) for hotter stars.
Figure~\ref{comp_lamost_hrs} also shows that $\Delta\log{g}\,{\rm (LAMOST-HRS)}$
    tend to increase with decreasing $\log{g}$ for $\log{g}\lesssim2.5$, and
    with decreasing $T_{\rm eff}$ for $T_{\rm eff}<5,000$\,K, where the HR
    samples were mostly from giant stars studied by Th2012, for which $\log{g}$
    values were derived using the asteroseismology method.
Th2012 presented the stellar parameters based on pure spectroscopic methods as
    well.
In Figure~\ref{comp_thygesen} we plot the differences between the LAMOST
    $\log{g}$ and the asteroseismic and spectroscopic $\log{g}$ in Th2012, as
    functions of $T_{\rm eff}$.
It is obvious that, for both cases, the trends of $\Delta\log{g}$ are quite
    similar.
Moreover, previous studies (e.g. Th2012, \citealt{Takeda2015}) have shown that
    $\log{g}$ obtained by the two methods are satisfactorily similar for giants.
These facts suggest that LASP overestimated $\log{g}$ by up to 0.5\,dex for cool
    giants.
For metallicity, the scatter tends to increase with decreasing $T_{\rm eff}$ and
    $\log{g}$.

\section{METHOD}
\subsection{Asteroseismic Analysis}
\label{method}
Solar-like oscillations are excited by the near-surface turbulent convection in
    a star, which is characterized by the global oscillation parameters
    $\Delta\nu$, corresponding to the average frequency separation between
    oscillation modes with consecutive radial orders $n$ and the same spherical
    degree $l$, and $\nu_{\rm max}$, the frequency at which the oscillation
    power is maximum.
The parameter $\Delta\nu$ is proportional to the square root of the mean stellar
    density ($\rho$) and is therefore given by \citep{Ulrich1986}:
    \begin{equation}\label{dnu}
        \Delta\nu = \sqrt{\frac{M/M_\odot}{(R/R_\odot)^3}}\Delta\nu_\odot
    \end{equation}
    with respect to the Sun.
The parameter $\nu_{\rm max}$ is assumed to be scaled with the acoustic cutoff
    frequency \citep{Brown1991}, and \citet{Kjeldsen1995} used this assumption
    to relate $\nu_{\rm max}$ to the fundamental stellar parameters as follows.
    \begin{equation}\label{numax}
        \nu_{\rm max} = \frac{M/M_\odot}{(R/R_\odot)^2\sqrt{T_{\rm eff}/T_{\rm
        eff, \odot}}}\nu_{\rm max,\odot}
    \end{equation}
By solving Equations~\ref{dnu} and \ref{numax}, one can obtain the relations
    linking the stellar mass $M$, radius $R$, mean density $\rho$, and surface
    gravity $\log{g}$ with the global oscillation parameters $\Delta\nu$ and
    $\nu_{\rm max}$.
It is noted that $\log{g}$ only depends on $\nu_{\rm max}$ for a given $T_{\rm
    eff}$.

The values of $\Delta\nu$ and $\nu_{\rm max}$ for different types of stars in
    the {\em Kepler} field have been used to estimate $M$, $R$, $\rho$ and
    $\log{g}$ in various studies.
For instance, \citet{Kallinger2010} determined the parameters for $>1,000$ red
    giants based on the first 138 days of the {\em Kepler} photometric data.
\citet{Hekker2011} used the data of the first 33 days to characterize more than
    10,000 giants for which solar-like oscillations have been detected.
This work was later refined by \cite{Stello2013} using the {\em Kepler} data
    with a longer time baseline of 681 days.
The {\em Kepler} mission also detected solar-like oscillations for 500 out of
    2,000 pre-selected main sequence and sub-giant stars during the first 10
    months of its scientific operation \citep{Chaplin2011}.
The fundamental parameters of these stars were published in \cite{Chaplin2014}
    and led to better characterization of planets \citep{Huber2013} and their
    host stars \citep{Mathur2011,Johnson2014}.
\citet{Huber2014} presented the revised catalog of parameters for more than
    190,000 stars for the {\em Kepler} Quarter 1-16 data.

Although stellar parameters can be directly derived Equations~\ref{dnu} and
    \ref{numax} as
\begin{align}
\left(\frac{M}{M_\odot}\right) & = \left(\frac{\nu_\mathrm{max}}{\nu_\mathrm{max,\odot}}\right)^3
                                   \left(\frac{\Delta\nu}{\Delta\nu_\odot}\right)^{-4}
                                   \left(\frac{T_\mathrm{eff}}{T_\mathrm{eff,\odot}}\right)^{3/2}
                                   \label{para1}
                                   \\
\left(\frac{R}{R_\odot}\right) & = \left(\frac{\nu_\mathrm{max}}{\nu_\mathrm{max,\odot}}\right)
                                   \left(\frac{\Delta\nu}{\Delta\nu_\odot}\right)^{-2}
                                   \left(\frac{T_\mathrm{eff}}{T_\mathrm{eff,\odot}}\right)^{1/2}
                                   \label{para2}
                                   \\
\left(\frac{\rho}{\rho_\odot}\right) & = \left(\frac{\Delta\nu}{\Delta\nu_\odot}\right)^2
                                   \label{para3}
            \\
\log{g} & = \log{g_\odot} + \log\left({\frac{\nu_\mathrm{max}}{\nu_\mathrm{max,\odot}}}\right)
            + \frac12\log\left({\frac{T_\mathrm{eff}}{T_\mathrm{eff,\odot}}}\right)
                                   \label{para4}
            \\
\left(\frac{L}{L_\odot}\right) & = \left(\frac{\nu_{\rm max}}{\nu_{\rm max,\odot}}\right)^2
                                   \left(\frac{\Delta\nu}{\Delta\nu_\odot}\right)^{-4}
                                   \left(\frac{T_{\rm eff}}{T_{\rm eff,\odot}}\right)^5
                                   \label{para5}
\end{align},
    some sets of ($M$, $R$, $T_{\rm eff}$) for a given metallicity are not
    permitted according to the stellar evolution theories.
Grid-based methods containing a significantly large number of parameters ($M$,
    $R$, $T_{\rm eff}$, [Fe/H]) returned by stellar evolution programs have been
    widely used to find the best match to the observed parameters
    \citep[see][and references therein]{Chaplin2013}.
We adopted the Geneva stellar evolutionary tracks \citep{Lejeune2001}, which
    cover a wide range of mass and metallicity ($Z$) values.
The values of high-temperature opacities were taken from the OPAL data
    \citep{Iglesias1996}, and those of low-temperature opacities were taken from
    \citet{Kurucz1991} or \citet{Alexander1994}.
For stars with $M\le1.5M_\odot$, a core overshooting parameter of
    $d/H_{\rm P}=0.2$ was adopted.
Mass loss of \citet{Reimers1975} and \citet{Jager1988} were taken into account.
In previous grid-based analyses \citep[e.g.][]{Basu2010, Kallinger2010,
    Huber2014}, some widely used stellar models, such as the YREC \citep[Yale
    Stellar Evolution Code;][]{Demarque2008}, DSEP \citep[Dartmouth Stellar
    Evolution Program;][]{Dotter2008}, and BaSTI \citep[Bag of Stellar Tracks
    and Isochrones;][]{Piersanti2004} models, did not account for the
    evolutionary stages after the helium flash.
On the contrary, the evolution phases of Geneva database were calculated to the
    end of the early asymptotic giant branch (EAGB) phase for intermediate-mass
    stars ($2 \le M/M_\odot \le 5$), and to the end of the carbon burning phase
    for larger mass stars ($M/M_\odot \ge 7$).
Therefore, the evolutionary stages following the helium flash were included for
    stars with $M > 2 M_\odot$.
As a substantial number of our samples met the above condition, we considered
    that our method naturally eliminates the systematic bias towards larger
    masses for giants in \cite{Huber2014}, where post-helium flash data were not
    included for calculations.

To ensure that at least $10^2$ models are available for the final probability
    density function (PDF) of each star, we generated a dense grid by
    interpolating the evolutionary tracks in steps of 0.02\,dex for [Fe/H],
    ranging from $-2.0$ to +1.5, and steps of 0.02\,$M_\odot$ for the initial
    mass ($M_0$), ranging from 0.8 to 5.0\,$M_\odot$.
For each track, the Geneva database contained at most 51 groups of data points
    with $T_{\rm eff}$, $L$, age, and $M$.
Here, $M$ is the stellar mass, varying with time due to the mass loss.
We interpolated 500 points along the entire time span, and calculated $R$ and
    $\log{g}$ using the basic physical relations, along with $\Delta\nu$ and
    $\nu_{\rm max}$ that were calculated according to the scaling relations in
    Equations~\ref{dnu} and \ref{numax}, for each interpolated point.
We adopted the solar seismic parameters $\Delta\nu_\odot=135.1\pm0.1\,\mu$Hz and
    $\nu_{{\rm max},\odot}=3,090\pm30\,\mu$Hz that were based on the data
    collected by VIRGO aboard SOHO spacecraft during $\sim$11,000 days
    \citep{Huber2011}.
Our complete grid had a total of $\sim1.8\times10^7$ points, each containing
    nine parameters, $T_{\rm eff}$, $Z$, $M$, $R$, $L$, $\log{g}$, age,
    $\Delta\nu$, and $\nu_{\rm max}$.

Stellar fundamental parameters can be subsequently derived from the observed
    oscillation parameters ($\Delta\nu$ and $\nu_{\rm max}$) using the Bayesian
    approach, if $T_{\rm eff}$ and [Fe/H] are known.
The Bayes' theorem can be stated as
    \begin{equation}\label{bayesian}
        p(\boldsymbol\theta|\boldsymbol{d},M) = \frac{p(\boldsymbol\theta|M)
        p(\boldsymbol{d}|\boldsymbol\theta,M)}{p(\boldsymbol{d}|M)}
    \end{equation},
    where $p(\boldsymbol\theta|\boldsymbol{d},M)$ is the posterior probability
    distribution of parameters $\boldsymbol\theta$ for a certain model $M$,
    based on the observational data $\boldsymbol{d}$.
The model $M$ stands for an individual datum corresponding to an evolutionary
    status in our grid.
The distribution $p(\boldsymbol\theta|M)$ is the prior probability distribution
    of $\boldsymbol\theta$, and the likelihood function
    $p(\boldsymbol{d}|\boldsymbol\theta,M)$ is the probability of obtaining
    $\boldsymbol{d}$, given the parameters $\boldsymbol\theta$ for model $M$.
The quantity $1/p(\boldsymbol{d}|M)$ is the normalization term.
In our case, the observational data set is $\boldsymbol{d}=(T_{\rm eff},
    {\rm [Fe/H]}, \Delta\nu, \nu_{\rm max})$, and
    \begin{equation}
        p(\boldsymbol{d}|\boldsymbol\theta,M) = 
        {\cal L}_{T_{\rm eff}}
        {\cal L}_{\rm [Fe/H]}
        {\cal L}_{\Delta\nu}
        {\cal L}_{\nu_{\rm max}}
    \end{equation}.
The likelihood functions of each parameter are calculated to match the
    observational ones by assuming independent Gaussian-distributed errors.
Therefore, we have
    \begin{equation}
    {\cal L}_d = \frac{1}{\sqrt{2\pi}\sigma_d}
    \exp\left[-\frac{(d_{\rm obs}-d_{\rm model})^2}{2\sigma_d^2}\right]
    \end{equation}.

Some previous studies adopted uniform priors $p(\boldsymbol\theta|M)$ for all
    models in the grid \citep[e.g.][]{Kallinger2010}.
However, it should be noted that for a star with given ($M_0, Z$), the
    probability of its physical quantities being ($T_{\rm eff}, R, L, \log{g}$)
    when the star is being observed is inversely proportional to the star's
    evolutionary speed in its current stage.
Otherwise, the resulting stellar parameters would be biased towards the rapid
    evolution phases (see the description of the GOE pipeline in \citealt{
    Chaplin2014}).
In our approach, the differential age of a track with a given ($M_0, Z$) can
    well represent the reciprocals of the evolutionary speeds; thus
    \begin{equation}\label{pnorm}
        p(\boldsymbol\theta|M_{i,j}) = C
        \frac{a_{i+1,j}-a_{i,j}}{a_{n,j}-a_{1,j}}
        \quad i=1,2,\cdots n-1
    \end{equation},
    where $a_{i,j}$ is the age of the $i$-th interpolated point in the $j$-th
    track, $C$ denotes the normalization factor, and $n=500$ is the number of
    interpolated points along each track.
In the above equation, the time span of two adjacent points
    ($a_{i+1,j}-a_{i,j}$) is normalized by the total time ($a_{n,j}-a_{1,j}$)
    of the $j$-th track; otherwise, the posterior probability distributions
    would be biased towards low-mass stars.
Although larger-mass stars have shorter lifetimes than less massive stars, and
    hence, have lower probabilities of being observed as their higher
    luminosities make them visible over longer distances to a magnitude-limited
    survey, which, to some extent, cancels out the above age selection effect.
Therefore, aim of Equation~\ref{pnorm} only corrects the bias caused by
    different evolutionary speeds at different stages, rather than lifetimes, as
    a function of the stellar masses.
In our study, uniform probabilities for stars with different ($M_0$, $Z$) were
    assumed, because our observed data $\boldsymbol{d}$ were accurate, and the
    prior probabilities of ($M_0$, $Z$) were not expected to vary significantly
    over such a relatively narrow parametric range.

All the sample stars in this work have been monitored by the {\em Kepler} space
    telescope with extremely high photometric precision during its scientific
    operation.
Several research groups have devoted attention to extracting the values of
    $\Delta\nu$ and $\nu_{\rm max}$ from the {\em Kepler} light curves using
    various techniques \citep[e.g.,][]{Mosser2009, Huber2009, Kallinger2010a,
    Hekker2010}.
We employed the parameters from different literature sources, as listed in
    Table~\ref{tab-oscref}.
For nonseismic parameters $T_{\rm eff}$ and [Fe/H], we used the values returned
    by the LASP in the LAMOST AFGK-type star parameters catalog.

\begin{table*}
    \caption{
        Sources of oscillation parameters ($\Delta\nu$ and $\nu_{\rm max}$) for
        stars analyzed in this work. Methods for obtaining the seismic data are
        described by \citealt{Kallinger2010a} (CAN), \citealt{Hekker2010} (OCT),
        \citealt{Mosser2009} (COR), and \citealt{Huber2009} (SYD).
        }\label{tab-oscref}
  \begin{center}
    \begin{tabular}{lllcrr}
      \hline
      \hline
      Reference & Type of stars & Timespan & Method & $N_{\rm total}$ & $N_{\rm adopt}$ \\
      \hline
      \citet{Kallinger2010}&red giants and clump stars       &$\sim$1200 days (Q1 $\sim$ Q13)& CAN &$>1000$     &  630 \\
      \citet{Hekker2011}   &red giants and clump stars       &33 days                        & OCT &$>10^4$     & 1548 \\
      \cite{Mathur2011}    &two solar-type stars             &8 months                       &  -- &      2     &    1 \\
      \citet{Mosser2012}   &red giants and clump stars       &690 days (Q1 $\sim$ Q8)        & COR &    218     &   10 \\
      \citet{Huber2013}    &planet-candidate host stars      &$\sim$1000 days (Q1 $\sim$ Q11)& SYD &     77     &   27 \\
      \citet{Stello2013}   &red giants and clump stars       &681 days (Q0 $\sim$ Q8)        & SYD &$\sim$13000 &  630 \\
      \citet{Chaplin2014}  &main-sequence and sub-giant stars&$\sim$300 days                 & SYD & 518        &  214 \\
      \hline                                                                                  
                           &                                 &                               & & Total          & 3060 \\
      \hline
    \end{tabular}
  \end{center}
\end{table*}

\subsection{Iterative Process}
\label{iterative}
In our work, the derived $\log{g}$ obtained using the above approach could
    differ from spectroscopically obtained values by as much as 0.5\,dex (see
    Section~\ref{lamost_calib}), which could in turn yield a significant bias in
    $T_{\rm eff}$ and [Fe/H].
Therefore, we determined our spectroscopic parameters ($T_{\rm eff}$, [Fe/H] and
    $\log{g}$) iteratively.
First, asteroseismic $\log{g}$ values (hereafter, $\log{g}_{\rm iter0}$) were
    obtained by using the above-mentioned grid method, with $T_{\rm eff}$ and
    [Fe/H] listed in the LAMOST catalog (hereafter, $T_{\rm eff, LASP}$, and
    [Fe/H]$_{\rm LASP}$), and oscillation parameters $\Delta\nu$ and
    $\nu_{\rm max}$.
Then, the LAMOST spectra for all the sample stars were reanalyzed by LASP with
    fixed $\log{g}_{\rm iter0}$, to acquire new $T_{\rm eff, iter1}$ and
    [Fe/H]$_{\rm iter1}$ values, which were then used for calculating
    asteroseismic $\log{g}_{\rm iter1}$.
We found that, in our sample, a change of +0.1\,dex in $\log{g}$ resulted in
    $\Delta T_{\rm eff}\sim$ +27\,K and $\Delta$[Fe/H] $\sim$ +0.02\,dex for
    giants, and in $\Delta T_{\rm eff}\sim$ +36\,K and $\Delta$[Fe/H] $\sim$
    +0.01\,dex for dwarfs.
The differences between $\log{g}_{\rm iter1}$ and $\log{g}_{\rm iter0}$ were
    within $\pm$0.03\,dex for 99\% of our giants, and $\pm$0.01\,dex for all of
    our dwarfs, except for only one star.
These small changes in $\log{g}$ after the first iteration had negligible
    effects on $T_{\rm eff}$ and [Fe/H] compared with the observational
    uncertainties because, according to Equation~\ref{para4}, asteroseismic
    $\log{g}$ only depends weakly on $T_{\rm eff}$.
Consequently, our results regarding atmospheric parameters converged after one
    iteration.

\section{RESULTS}

\subsection{Stellar Parameters}

\begin{figure*}
    \centering
    \includegraphics[width=16cm]{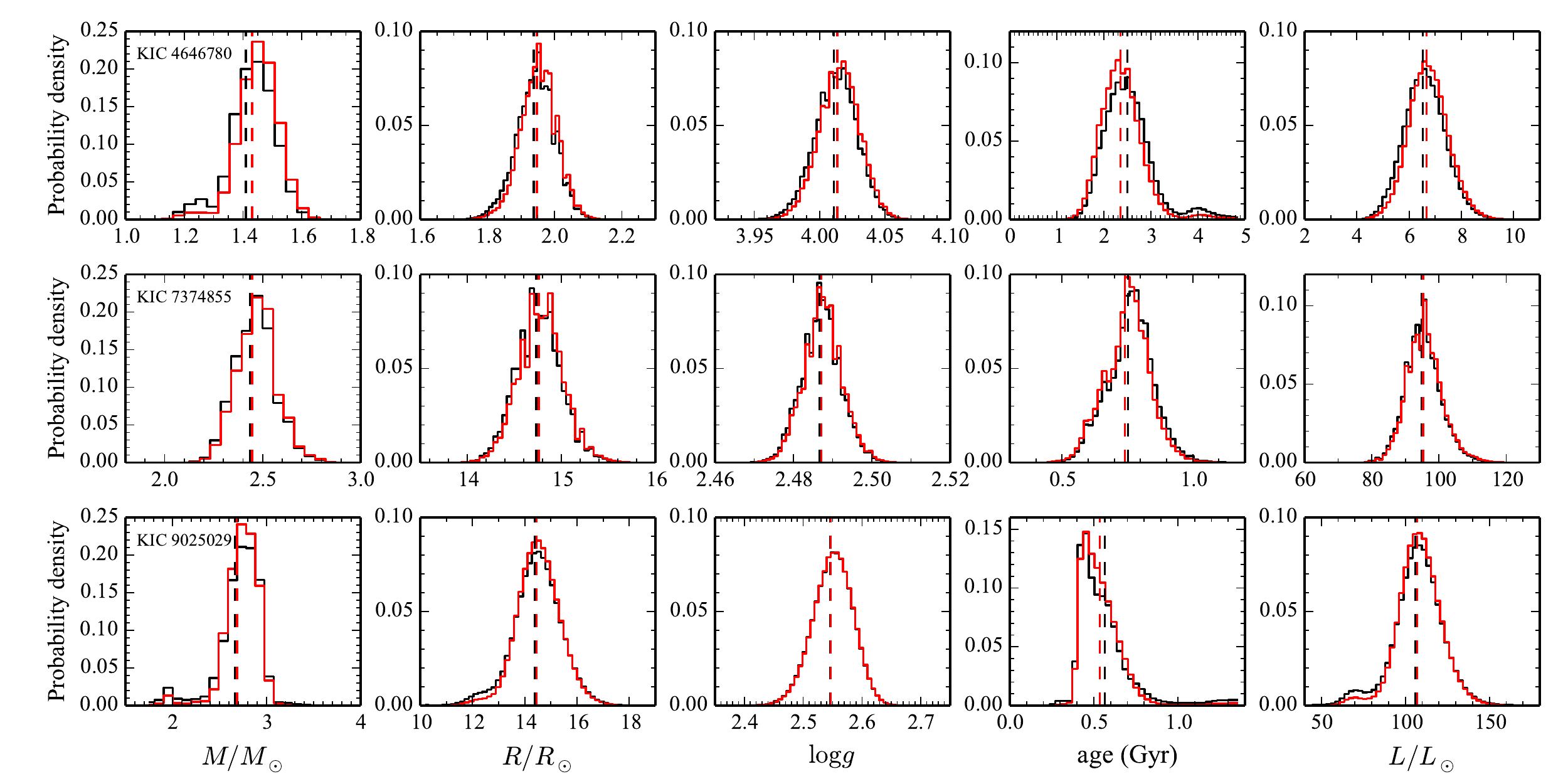}
    \caption{
        Examples of normalized PDFs of stellar parameters, for one dwarf
        (KIC\,4646780) and two giants (KIC\,7374855 and KIC\,9025029).
        Columns from left to right: $M$, $R$, $\log{g}$, age, and $L$.
        Red and black histograms indicate time-weighted and nonweighted PDFs.
        Vertical dashed lines represent the corresponding average values.
    }
    \label{fig_pdf}

\end{figure*}

\begin{table*}
    \caption{Stellar parameters of {\em Kepler} planet candidate hosts.}
    \label{tab-host}
    \footnotesize{
    \begin{center}
        \begin{tabular}{rrllccrrrr}
            \hline
            \hline
            KOI & KIC& Kepler name&$T_{\rm eff}$ (K) & $\log{g}$ & [Fe/H] & $M_\star/M_\odot$ & $R_\star/R_\odot$ & $L_\star/L_\odot$ & Age (Gyr) \\
            \hline
               5 &  8554498 &  & $5863\pm91$ & $4.015\pm0.013$ & $+0.130\pm0.110$ & $1.23\pm0.09$ & $1.80\pm0.05$ & $3.55\pm0.30$ & $6.10\pm2.44$ \\
   7 & 11853905 & Kepler-4 & $5820\pm93$ & $4.110\pm0.010$ & $+0.170\pm0.110$ & $1.15\pm0.07$ & $1.56\pm0.04$ & $2.62\pm0.22$ & $7.11\pm2.08$ \\
  41 &  6521045 & Kepler-100 & $5888\pm91$ & $4.130\pm0.008$ & $+0.090\pm0.110$ & $1.14\pm0.06$ & $1.52\pm0.04$ & $2.59\pm0.19$ & $7.23\pm1.88$ \\
  75 &  7199397 &  & $5913\pm88$ & $3.758\pm0.008$ & $-0.110\pm0.120$ & $1.36\pm0.05$ & $2.55\pm0.04$ & $7.09\pm0.54$ & $3.51\pm0.26$ \\
  85 &  5866724 & Kepler-65 & $6264\pm101$ & $4.232\pm0.006$ & $+0.130\pm0.110$ & $1.25\pm0.05$ & $1.42\pm0.02$ & $2.81\pm0.24$ & $3.20\pm0.97$ \\
 108 &  4914423 & Kepler-103 & $5901\pm98$ & $4.167\pm0.010$ & $+0.100\pm0.120$ & $1.14\pm0.06$ & $1.46\pm0.03$ & $2.38\pm0.19$ & $6.80\pm1.78$ \\
 122 &  8349582 & Kepler-95 & $5652\pm113$ & $4.168\pm0.012$ & $+0.220\pm0.150$ & $1.07\pm0.07$ & $1.41\pm0.03$ & $1.96\pm0.23$ & $8.42\pm2.15$ \\
 123 &  5094751 & Kepler-109 & $5992\pm107$ & $4.216\pm0.014$ & $-0.030\pm0.130$ & $1.09\pm0.06$ & $1.35\pm0.03$ & $2.15\pm0.20$ & $7.42\pm1.52$ \\
 262 & 11807274 & Kepler-50 & $6271\pm101$ & $4.132\pm0.008$ & $-0.090\pm0.120$ & $1.23\pm0.07$ & $1.58\pm0.03$ & $3.41\pm0.34$ & $4.30\pm1.49$ \\
 269 &  7670943 &  & $6481\pm133$ & $4.233\pm0.010$ & $-0.060\pm0.140$ & $1.29\pm0.06$ & $1.43\pm0.03$ & $3.28\pm0.36$ & $2.46\pm0.87$ \\
 271 &  9451706 & Kepler-127 & $6142\pm100$ & $4.261\pm0.010$ & $+0.250\pm0.110$ & $1.22\pm0.05$ & $1.35\pm0.02$ & $2.36\pm0.22$ & $3.55\pm1.15$ \\
 275 & 10586004 & Kepler-129 & $5802\pm93$ & $4.086\pm0.009$ & $+0.230\pm0.110$ & $1.23\pm0.06$ & $1.66\pm0.04$ & $2.93\pm0.21$ & $5.33\pm1.46$ \\
 276 & 11133306 &  & $6013\pm99$ & $4.322\pm0.010$ & $-0.030\pm0.120$ & $1.08\pm0.05$ & $1.19\pm0.02$ & $1.70\pm0.16$ & $5.93\pm1.41$ \\
 280 &  4141376 &  & $6171\pm132$ & $4.420\pm0.009$ & $-0.320\pm0.160$ & $1.07\pm0.06$ & $1.05\pm0.02$ & $1.48\pm0.15$ & $2.84\pm1.81$ \\
 281 &  4143755 &  & $5699\pm118$ & $4.108\pm0.010$ & $-0.480\pm0.180$ & $0.97\pm0.04$ & $1.43\pm0.03$ & $2.09\pm0.21$ & $10.18\pm0.91$ \\
 285 &  6196457 & Kepler-92 & $5908\pm89$ & $4.059\pm0.010$ & $+0.150\pm0.110$ & $1.23\pm0.07$ & $1.71\pm0.04$ & $3.26\pm0.24$ & $5.56\pm1.85$ \\
 319 &  8684730 &  & $5879\pm88$ & $3.930\pm0.015$ & $+0.080\pm0.110$ & $1.29\pm0.12$ & $2.04\pm0.09$ & $4.66\pm0.58$ & $5.51\pm2.57$ \\
 370 &  8494142 & Kepler-145 & $6102\pm102$ & $4.016\pm0.012$ & $+0.030\pm0.130$ & $1.25\pm0.09$ & $1.82\pm0.05$ & $4.08\pm0.38$ & $5.38\pm2.07$ \\
 623 & 12068975 & Kepler-197 & $6129\pm104$ & $4.315\pm0.014$ & $-0.420\pm0.140$ & $1.00\pm0.06$ & $1.15\pm0.03$ & $1.73\pm0.17$ & $6.67\pm1.91$ \\
 674 &  7277317 &  & $4750\pm107$ & $3.636\pm0.011$ & $+0.100\pm0.140$ & $1.10\pm0.10$ & $2.64\pm0.10$ & $3.67\pm0.44$ & $9.28\pm2.57$ \\
1221 &  3640905 & Kepler-278 & $4891\pm84$ & $3.612\pm0.006$ & $+0.270\pm0.110$ & $1.27\pm0.06$ & $2.91\pm0.05$ & $4.63\pm0.28$ & $5.37\pm1.12$ \\
1621 &  5561278 &  & $6115\pm107$ & $3.969\pm0.012$ & $-0.010\pm0.130$ & $1.32\pm0.10$ & $1.97\pm0.07$ & $4.83\pm0.55$ & $4.40\pm1.90$ \\
1925 &  9955598 & Kepler-409 & $5421\pm93$ & $4.496\pm0.005$ & $+0.130\pm0.130$ & $0.91\pm0.03$ & $0.89\pm0.01$ & $0.64\pm0.05$ & $7.39\pm1.83$ \\
2133 &  8219268 & Kepler-91 & $4464\pm80$ & $2.933\pm0.010$ & $+0.270\pm0.090$ & $1.32\pm0.09$ & $6.47\pm0.19$ & $17.56\pm1.39$ & $4.75\pm1.30$ \\
2706 &  9697131 &  & $6313\pm141$ & $4.022\pm0.013$ & $+0.120\pm0.140$ & $1.42\pm0.08$ & $1.92\pm0.05$ & $5.26\pm0.66$ & $2.79\pm1.13$ \\
2792 & 11127479 &  & $5982\pm97$ & $4.229\pm0.012$ & $+0.210\pm0.110$ & $1.15\pm0.06$ & $1.36\pm0.03$ & $2.13\pm0.21$ & $5.69\pm1.68$ \\
5110 &  4953262 &  & $4899\pm104$ & $3.186\pm0.036$ & $-0.130\pm0.140$ & $1.38\pm0.23$ & $4.94\pm0.44$ & $12.58\pm2.68$ & $4.32\pm2.81$ \\
5174 &  5524229 &  & $4971\pm87$ & $3.102\pm0.035$ & $-0.400\pm0.130$ & $1.41\pm0.23$ & $5.50\pm0.49$ & $16.33\pm3.45$ & $3.39\pm1.88$ \\
5223 &  6023571 &  & $4610\pm105$ & $2.984\pm0.020$ & $+0.180\pm0.120$ & $1.26\pm0.17$ & $5.96\pm0.32$ & $15.59\pm2.01$ & $6.12\pm3.08$ \\
5281 &  6425377 &  & $4941\pm148$ & $3.107\pm0.013$ & $-0.010\pm0.200$ & $1.27\pm0.15$ & $5.20\pm0.26$ & $13.11\pm1.59$ & $5.23\pm2.57$ \\
5287 &  6509282 &  & $4706\pm133$ & $2.528\pm0.035$ & $-0.110\pm0.160$ & $2.50\pm0.42$ & $14.20\pm1.34$ & $96.87\pm22.13$ & $0.81\pm0.74$ \\
5322 &  6756369 &  & $4808\pm114$ & $3.228\pm0.012$ & $-0.190\pm0.150$ & $1.26\pm0.13$ & $4.51\pm0.19$ & $10.31\pm1.13$ & $5.04\pm2.15$ \\
5543 &  8560475 &  & $4798\pm108$ & $2.448\pm0.023$ & $-0.160\pm0.130$ & $2.23\pm0.43$ & $14.67\pm1.08$ & $98.49\pm23.31$ & $1.02\pm0.52$ \\
5578 &  8868481 &  & $5582\pm90$ & $3.774\pm0.014$ & $-0.040\pm0.130$ & $1.26\pm0.07$ & $2.41\pm0.08$ & $5.05\pm0.51$ & $5.25\pm1.74$ \\
5782 & 10272858 &  & $4992\pm76$ & $3.448\pm0.009$ & $+0.020\pm0.100$ & $1.34\pm0.14$ & $3.61\pm0.18$ & $7.06\pm0.89$ & $4.46\pm2.09$ \\
6877 &  7431665 &  & $4661\pm82$ & $2.645\pm0.025$ & $-0.090\pm0.100$ & $1.75\pm0.29$ & $10.35\pm0.64$ & $47.06\pm8.90$ & $2.09\pm1.33$ \\
7296 & 10214328 &  & $6624\pm141$ & $4.017\pm0.014$ & $-0.350\pm0.140$ & $1.36\pm0.08$ & $1.89\pm0.05$ & $6.16\pm0.74$ & $2.77\pm0.61$ \\

            \hline
        \end{tabular}
    \end{center}
}
\end{table*}

\begin{table*}
    \caption{Stellar parameters for all the stars in this study.
             Only the first five rows are shown here to illustrate the format.
             The full table is available online.}
    \label{tab_allstars}
    \scriptsize{
    \begin{center}
        \begin{tabular}{llllllllrrrr}
            \hline
            \hline
            KIC & Kepler & KOI &S/N& Kp &
            $T_{\rm eff}$ & $\log{g}$ & [Fe/H] &
            Mass  & Radius  & Luminosity  &Age\\
                &        &     &   &   &
            (K)          &      &    &
            ($M_\odot$) & ($R_\odot$) & ($L_\odot$) & (Gyr)\\
            \hline
             1162746&   0&   0& 69&11.446&$ 4787 \pm 104 $&$ 2.359 \pm 0.002 $&$ -0.500 \pm 0.140 $&$ 0.880 \pm 0.005 $&$ 10.26 \pm  0.04 $&$  48.25 \pm   1.00 $&$ 11.36 \pm  0.39 $\\
 1163621&   0&   0& 35&11.791&$ 4942 \pm 120 $&$ 2.659 \pm 0.014 $&$ +0.020 \pm 0.150 $&$ 2.400 \pm 0.223 $&$ 11.98 \pm  0.49 $&$  71.67 \pm   8.38 $&$  0.79 \pm  0.27 $\\
 1294122&   0&   0& 47&11.938&$ 4761 \pm 100 $&$ 2.808 \pm 0.018 $&$ -0.040 \pm 0.120 $&$ 1.874 \pm 0.239 $&$  8.90 \pm  0.48 $&$  38.76 \pm   6.40 $&$  1.66 \pm  1.05 $\\
 1572049&   0&   0& 66&11.571&$ 5093 \pm 128 $&$ 2.679 \pm 0.014 $&$ -0.220 \pm 0.170 $&$ 2.450 \pm 0.141 $&$ 11.84 \pm  0.32 $&$  77.01 \pm   6.12 $&$  0.66 \pm  0.15 $\\
 1717618&   0&   0& 72&11.702&$ 5019 \pm  94 $&$ 3.003 \pm 0.016 $&$ -0.000 \pm 0.120 $&$ 1.910 \pm 0.186 $&$  7.19 \pm  0.26 $&$  27.09 \pm   3.57 $&$  1.36 \pm  0.46 $\\

            \hline
        \end{tabular}
    \end{center}
    }
\end{table*}

We applied the grid-based method, described in Section~\ref{method}, to derive
    the PDFs of $M$, $R$, $\log{g}$, $L$, and age, for 3,060 stars with SNR
    $>30$ spectra in the LAMOST-DR2 and DR3 Quarter 1 catalog.
For each PDF, we report its mean as the result, and use standard deviation as a
    measure of uncertainty.
Figure~\ref{fig_pdf} shows examples of PDFs of $M$, $R$, $\log{g}$, age and $L$
    for one typical main-sequence star and two evolved stars.
For comparison, we plot the time-weighted and non-weighted PDFs by using red and
    black solid curves, respectively.
These results show that by taking into account the evolution speed effect as
    discussed in Section~\ref{method}, the values of $M$, $R$, and $L$ shift
    towards higher values whereas the resulting age becomes smaller.
This is expected because the weights of the phases are reduced after evolving
    off the main sequence.

Stellar properties of planet candidate hosts are of particular interest because
    they are directly related to the planetary radii and masses in transit and
    Doppler detections.
Serious uncertainties in metallicities, surface gravities, and radii, mostly
    based on broad-band photometry, have been found in the KIC \citep[e.g.][]{
    Verner2011, Dong2014}, while high-resolution spectra are expensive for most
    of the {\em Kepler} planet hosts with $K_{\rm p}<13$ \citep[e.g.][]{
    Marcy2014}.
Alternatively, asteroseismology with spectroscopic inputs has been used for
    characterizing these planetary systems \citep[e.g.][]{Huber2013,
    Chaplin2013}.
There were 60 KOIs in our catalog, including 15 confirmed planet-host stars, 23
    ``false positives," and 22 host candidates awaiting validation.
In Table~\ref{tab-host} we list the results for the confirmed and candidate
    hosts.
The entire sample is available via an online catalog, and the first five rows
    are shown in Table~\ref{tab_allstars} to illustrate the format.

\begin{figure*}[htbp]
    \centering
    \includegraphics[width=15cm]{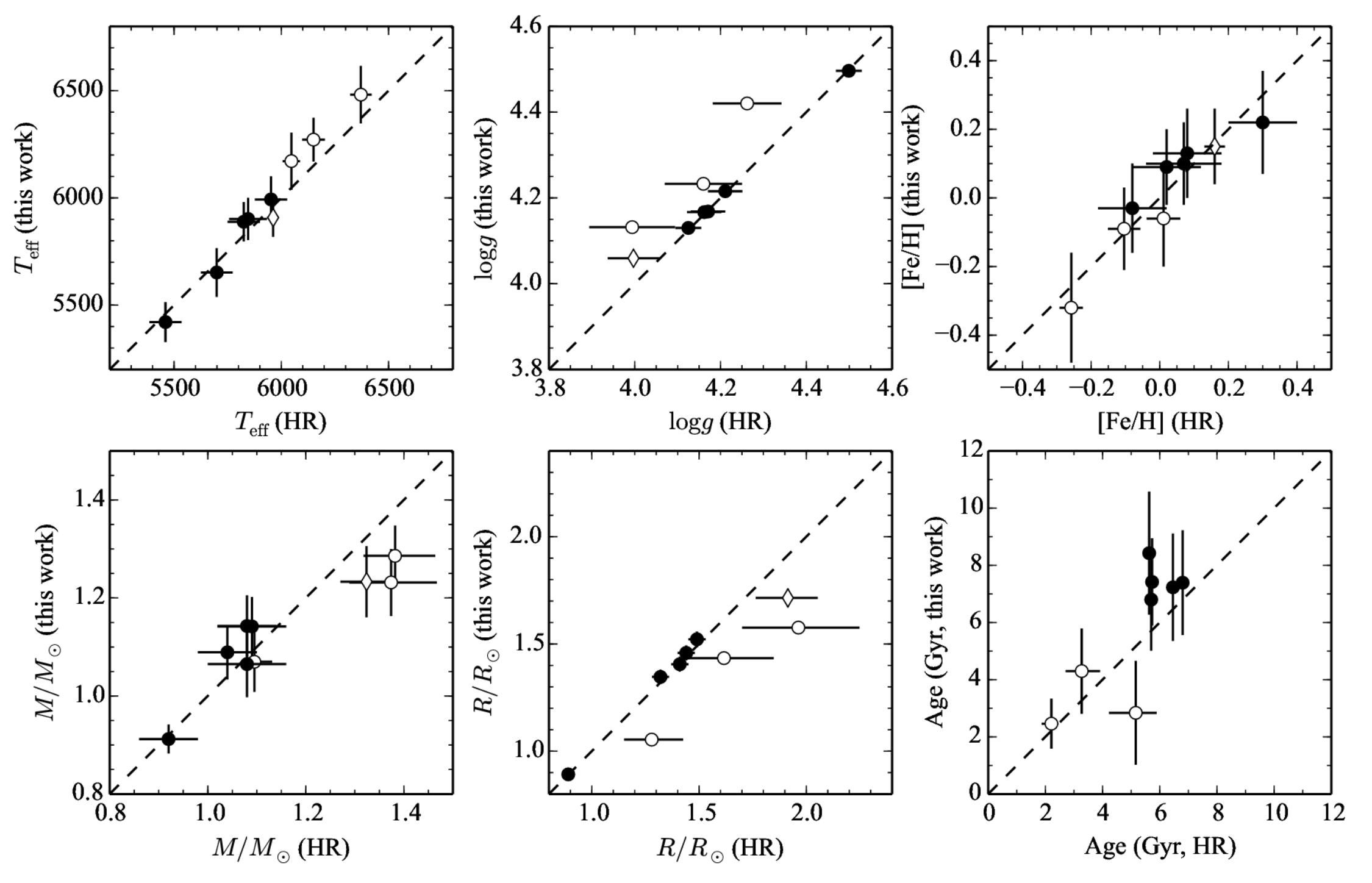}
    \caption{
        Comparison of stellar parameters for the KOIs in common with the
        previous studies that used high-resolution spectroscopy.
        Solid dots, open circles, and open diamonds represent parameters from
        M2014, Hi2012, and Hi2014, respectively.
    }
    \label{fig_compare_koi}
\end{figure*}

Figure~\ref{fig_compare_koi} compares the stellar parameters of the KOIs
    obtained in our work with those obtained in the previous studies that
    employed high-resolution spectroscopy.
There are five stars in common with M2014, and all of them show good agreements
    in terms of $T_{\rm eff}$, $\log{g}$, [Fe/H], $M$, and $R$.
Our derived age values were systematically higher than those in M2014, which is
    likely owing to the different theoretical evolution tracks used in these two
    studies (Y$^2$ in M14, and Geneva model in our study).
We also analyzed four common KOIs with Hi2012 and Hi2014, for which, the values
    of $T_{\rm eff}$ and [Fe/H] obtained were systematically lower while the
    values of $M$ and $R$ were higher than the previously reported results.
The stellar ages of two KOIs (KOI-269 and KOI-262) agreed within the
    corresponding error ranges, while the age of KOI-280 determined by us was
    lower.
Because the main-sequence stars evolve slowly on the HR diagram compared with
    the post-main sequence phases, age estimation by fitting the isochrones or
    evolution tracks is difficult and model-dependent (see \citealt{
    Soderblom2010}, and references therein).
Our estimation based on global oscillation parameters remains meaningful because
    the stellar ages are further constrained by $\Delta\nu$ and $\nu_{\rm max}$
    in addition to $T_{\rm eff}$, $\log{g}$, and [Fe/H].
A more detailed approach involves spectral analysis of excited oscillation
    modes \citep[e.g.][]{SilvaAguirre2015}.
Moreover, the field of gyrochronology, which has been developing with the help
    of the {\em Kepler} data, has made remarkable progress in refining the
    empirical relation between the stellar age and rotational period
    \citep[e.g.][]{Garcia2014, Angus2015}.

\subsection{Comparison with Huber et al. 2014}

\begin{figure*}
    \centering
    \includegraphics[width=6.9cm]{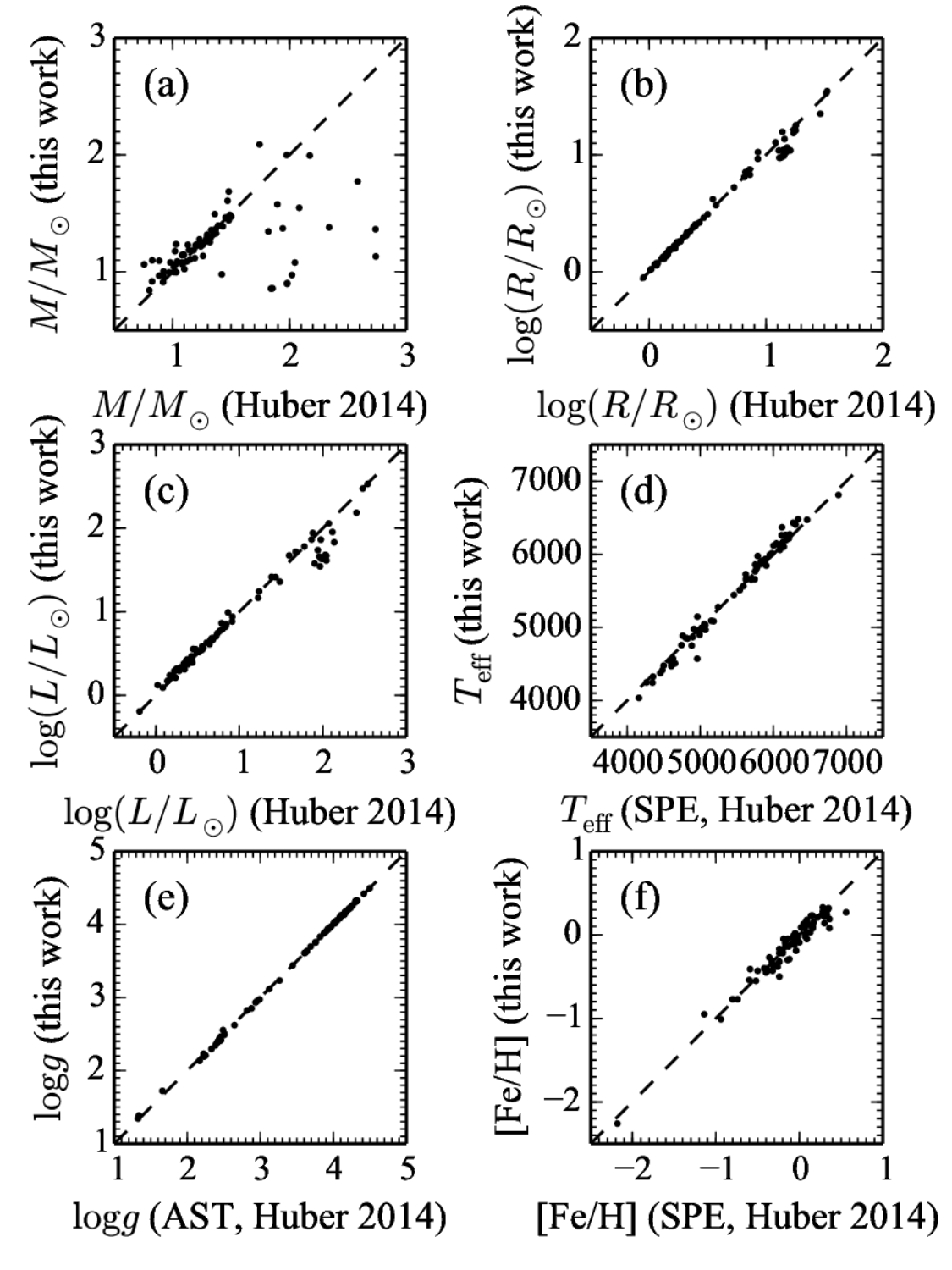}
    \includegraphics[width=9.2cm]{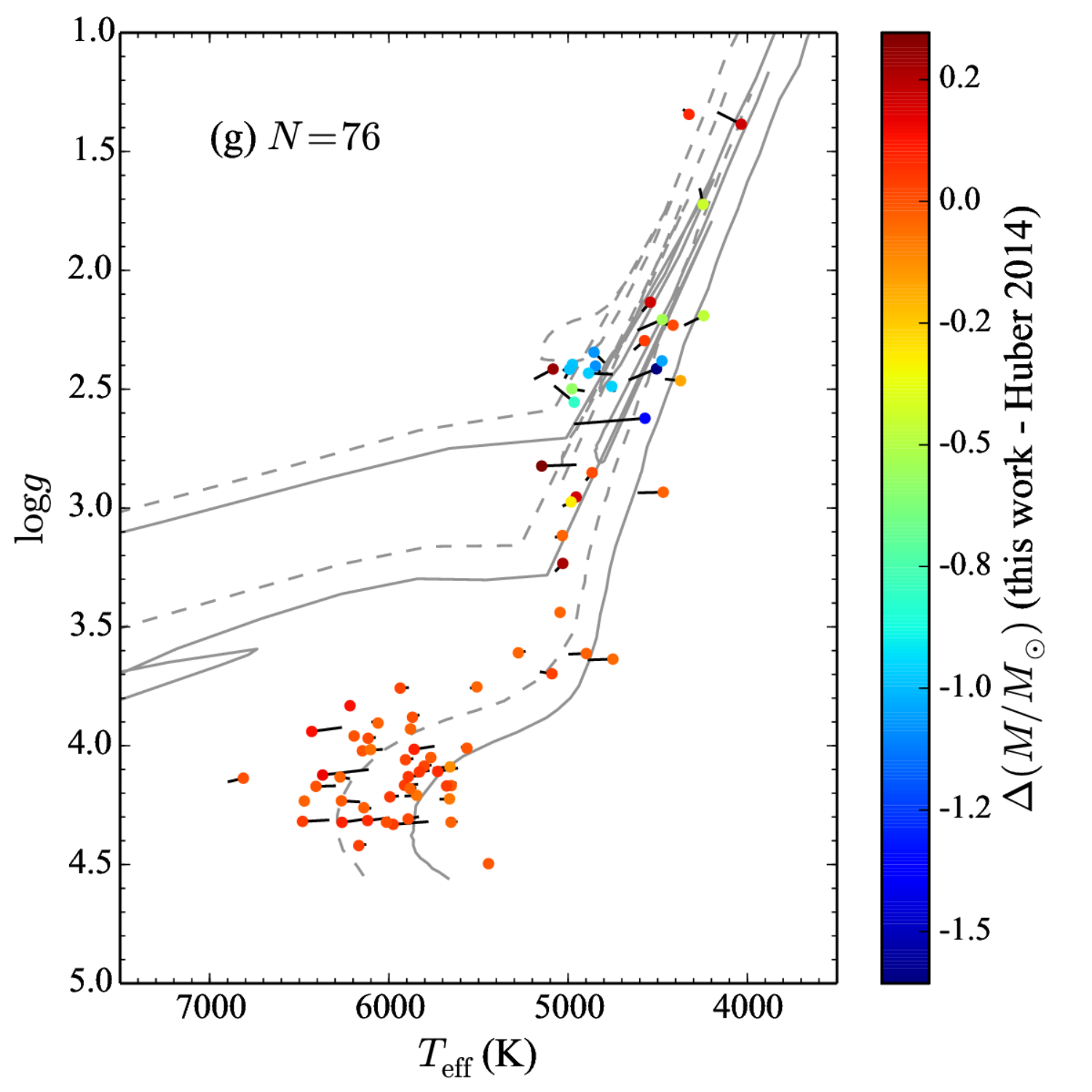}
    \caption{
        Comparisons of stellar parameters obtained in this work with those
        obtained from H2014 sub-category C.1 (see their Table~1).
        The values of $T_{\rm eff}$, $\log{g}$, and [Fe/H] in the H2014 study
        were obtained using spectroscopy, asteroseismology, and spectroscopy,
        respectively.
        Panel (g) shows the different positions of the sample stars on the Kiel
        diagram, by drawing solid lines to connect the parameters given by H2014
        with those obtained in the present work.
        The colors are coded with $\Delta M$ (this work $-$ H2014).
        A series of Geneva evolution tracks with initial masses $M_0 = 1.0$,
        2.0, and 3.0\,$M_\odot$, and [Fe/H] = 0.0 (solid lines), and $-0.5$
        (dashed lines) are also shown in panel (g).
    }
    \label{fig_huber_cat01}
\end{figure*}

\begin{figure*}
    \centering
    \includegraphics[width=6.9cm]{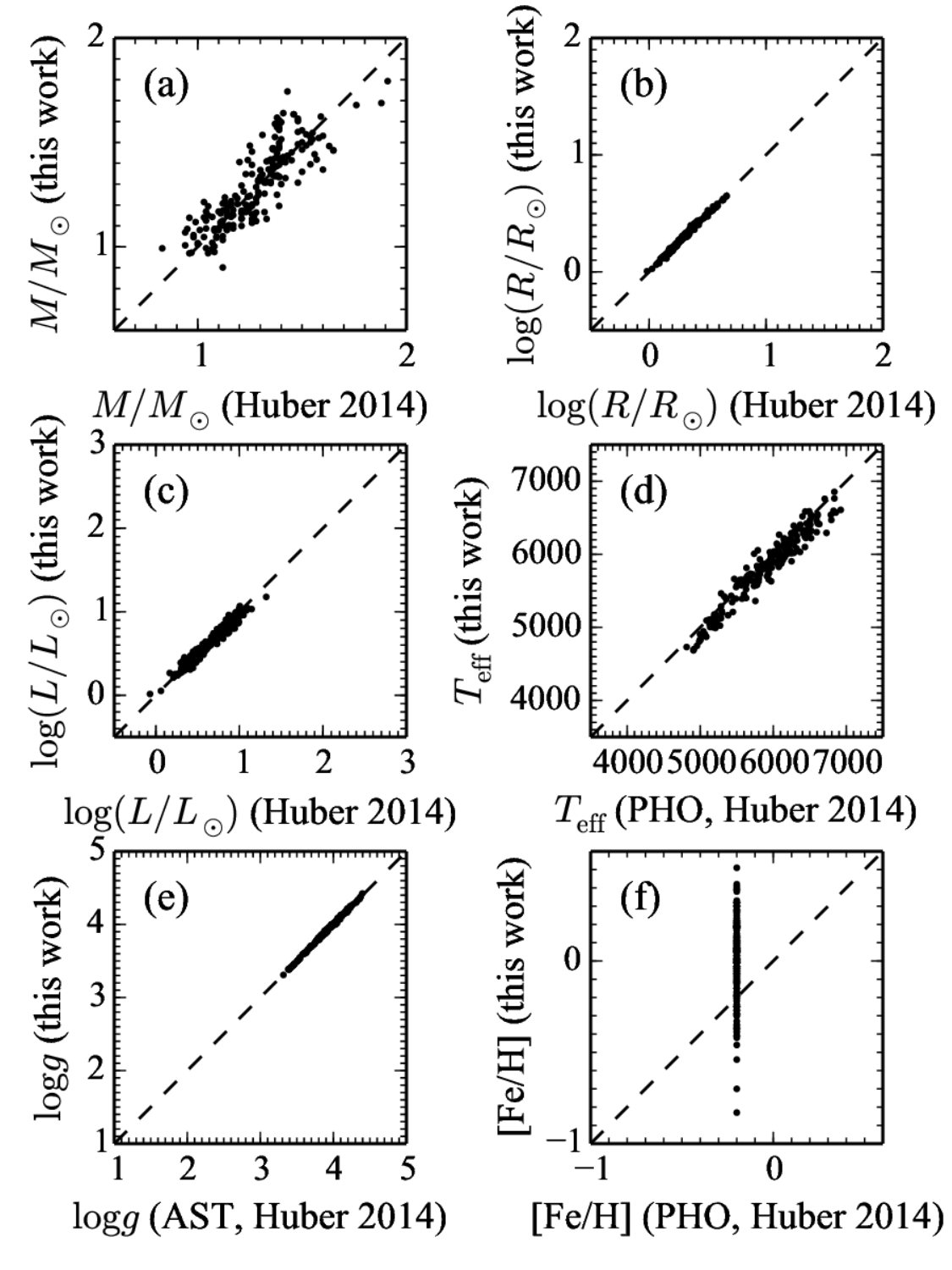}
    \includegraphics[width=9.2cm]{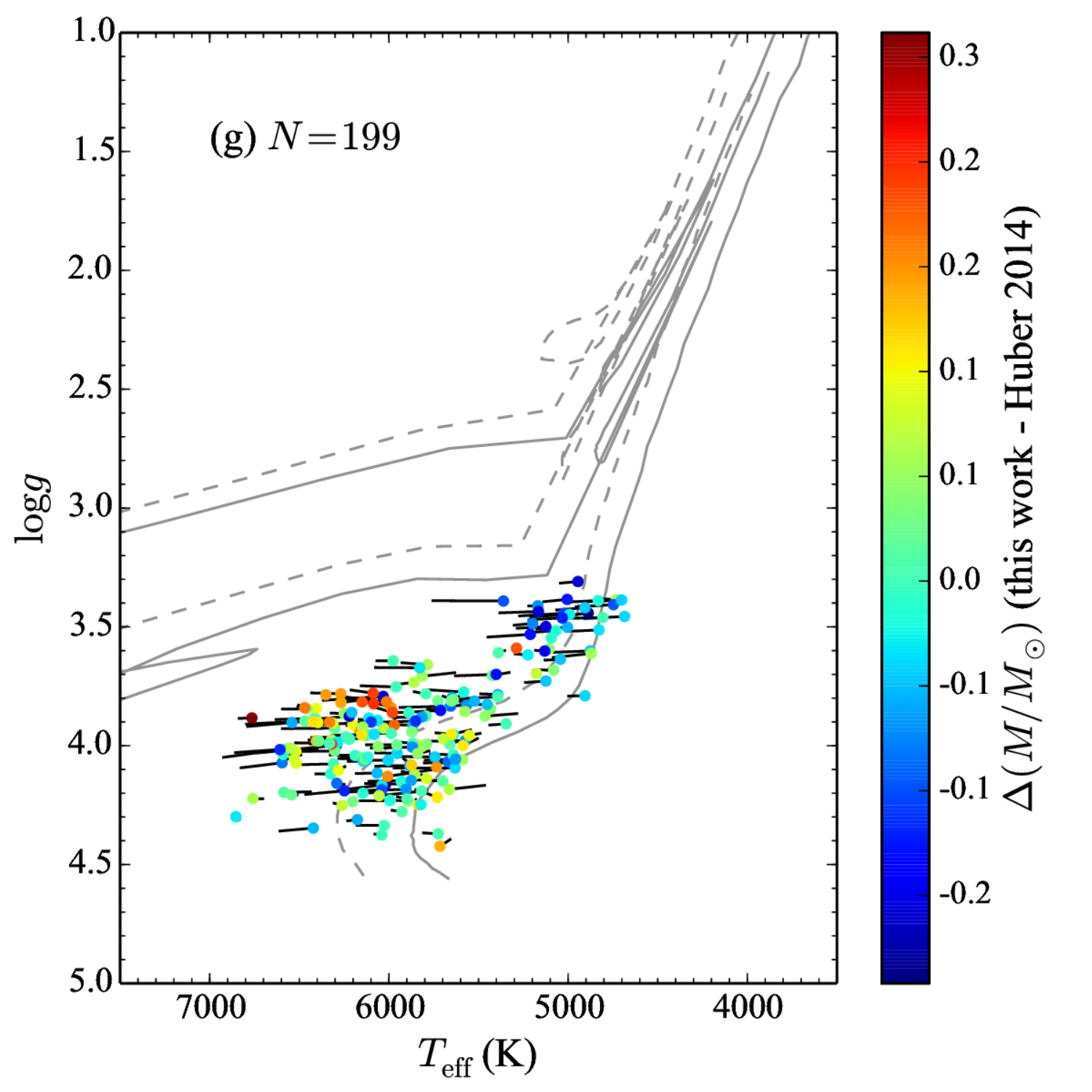}
    \caption{
        Same as Figure~\ref{fig_huber_cat01}, but for stars in common with
        sub-category C.4 in H2014, for which the values of $T_{\rm eff}$,
        $\log{g}$, and [Fe/H] were obtained using photometry, asteroseismology,
        and photometry, respectively.
    }
    \label{fig_huber_cat04}
\end{figure*}

\begin{figure*}
    \centering
    \includegraphics[width=6.9cm]{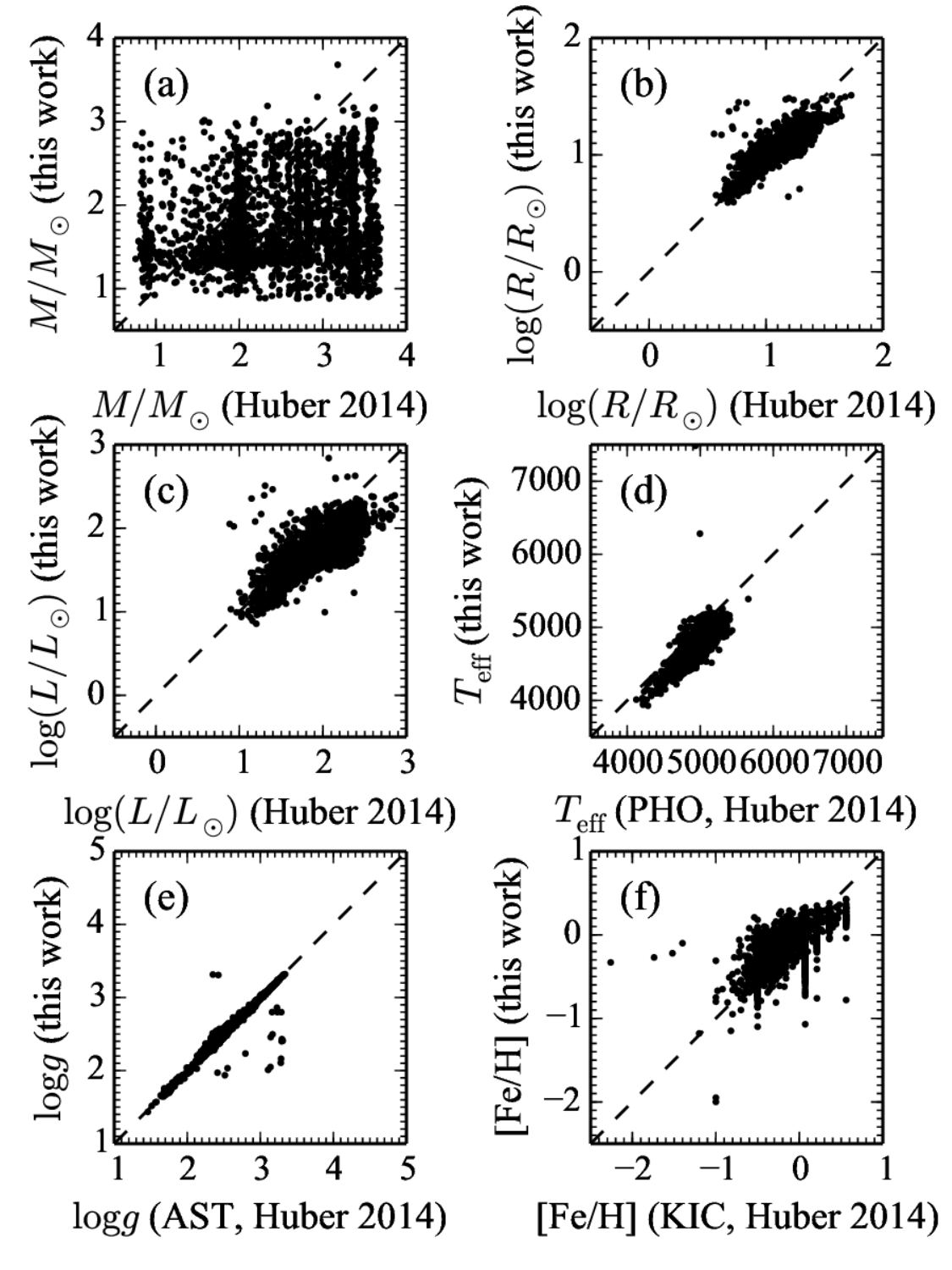}
    \includegraphics[width=9.2cm]{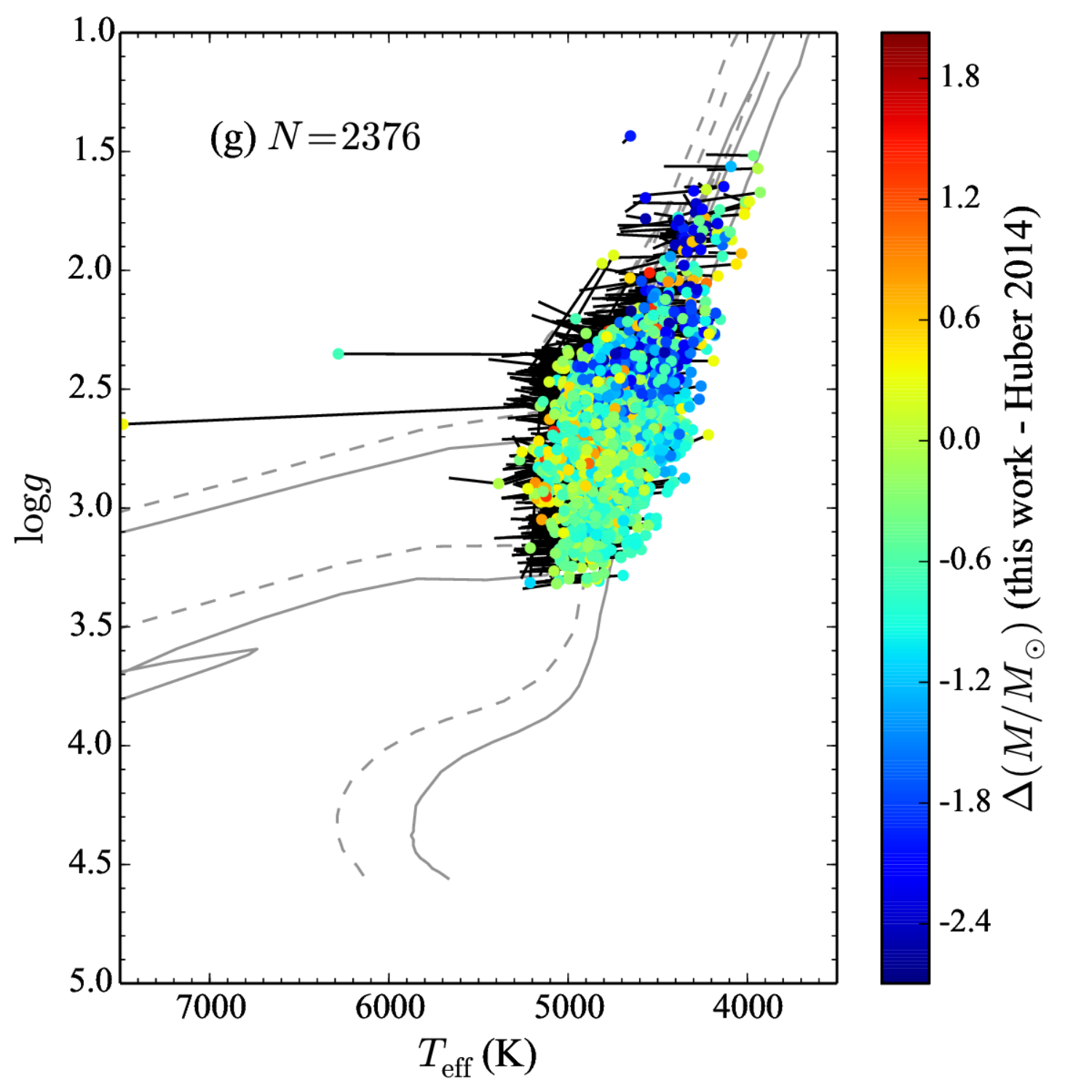}
    \caption{
        Same as Figures~\ref{fig_huber_cat01} and \ref{fig_huber_cat04}, but for
        stars in common with sub-category C.5 in H2014, for which the values of
        $T_{\rm eff}$, $\log{g}$, and [Fe/H] were obtained using photometry,
        asteroseismology, and the KIC, respectively.
    }
    \label{fig_huber_cat05}
\end{figure*}

Huber et al. (2014, hereafter, H2014) presented the stellar parameters for a
    large sample of {\em Kepler} stars observed in Quarter 1-16.
Their catalog is composed of several sub-categories designated by C.1-C.14 (see
    their Table~1), depending on the sources of input parameters ($T_{\rm eff}$,
    $\log{g}$, and [Fe/H]).
In this sub-section we focus on common stars within the H2014 sub-categories
    C.1, C.4, and C.5.
All of these three data sets adopted asteroseismic $\log{g}$, while
    $T_{\rm eff}$ and [Fe/H] were obtained using various techniques
    (spectroscopy, photometry, and KIC).
The category C.1 contains most of the ``gold-standard'' samples of H2014, for
    which high-resolution spectroscopy was used for the best possible
    characterization.
On the other hand, stars in C.4 and C.5 had no spectroscopic temperatures or
    metallicities.
In such cases, the authors of H2014 used a revised temperature scale by \cite{
    Pinsonneault2012}, and their [Fe/H] values were either fixed to $-0.2$ or
    obtained from KIC.

Figures~\ref{fig_huber_cat01} -- \ref{fig_huber_cat05} show the differences
    between the values of $M$, $R$, and $L$ (panels~a-c), as well as the
    atmospheric parameters $T_{\rm eff}$, $\log{g}$, and [Fe/H] (panels~d-f)
    that were used as inputs.
All of these reveal an excellent agreement between the values of $\log{g}$ in
    the two works, despite the fact that the values of $T_{\rm eff}$ and [Fe/H]
    were taken from different sources.
In each of the above figures, we also plot the differences between positions of
    the stars in the Kiel diagram, where the colored circles correspond to our
    results, and the other sides of solid lines correspond to the values from
    H2014.
The colors are coded with $\Delta M$ (this work $-$ H2014).
Geneva evolution tracks with $M=1.0$, 2.0, and 3.0\,$M_\odot$, and [Fe/H] = 0.0
    (solid gray lines) and $-0.5$ (dashed gray lines) are also shown.

Figure~\ref{fig_huber_cat01} shows the comparison of 76 common stars including
    dwarfs and giants, for which both H2014 and this work used spectroscopic
    $T_{\rm eff}$ and [Fe/H] as inputs for asteroseismic $\log{g}$ and other
    physical parameters.
Our results are in a good agreement with the previous work, with mean
    differences of only $0.00\pm0.02$, $-0.02\pm0.06$, and $-0.04\pm0.14$, for
    $\log{g}$, $\log{R}$, and $\log{L}$, respectively.
All outlying points in panel~(a) correspond to giants with $\log{g}<3.5$, for
    which evolution tracks are highly degenerated in the HR diagram.
We noted that our results on stellar masses for these giants are systematically
    lower than those from H2014.
This can be explained by the bias towards higher mass in the previous studies,
    as discussed in Section~\ref{method}.

Figure~\ref{fig_huber_cat04} compares the physical and spectroscopic parameters
    for 199 dwarfs and sub-giants in common with sub-category C.4 in H2014.
Although for all of the stars in this sub-category the values of [Fe/H] were
    fixed at $-0.2$, a good agreement between the two studies was found in terms
    of the values of $R$, $L$, and $\log{g}$.
On average, our $T_{\rm eff}$ values were $91\pm120$\,K lower than the
    previously reported values.
The mean difference between the stellar mass values was $0.01\pm0.10\,M_\odot$,
    as shown in panel~(a).

Figure~\ref{fig_huber_cat05} shows the same comparison, but for sub-category C.5
    in H2014.
In contrast to Figure~\ref{fig_huber_cat04}, the stars in C.5 are giants, with
    $\log{g}<3.5$.
Our $T_{\rm eff}$ values were $226\pm130$\,K lower than those reported in H2014,
    which subsequently significantly affected the stellar physical parameters.
In terms of $\log{g}$, our results were in good agreement with H2014, with the
    mean difference of only $0.00\pm0.08$\,dex.
The mean differences were $-0.03\pm0.23$\,dex for [Fe/H], $-0.07\pm0.10$ for
    $\log{R/R_\odot}$, and $-0.23\pm0.23$ for $\log{L/L_\odot}$.
Moreover, the derived stellar masses were generally lower than those in H2014,
    and the comparison of stellar masses in panel~(a) reveals a chaotic
    distribution.
We note that most of the masses in H2014 are in the 0.8-3.7\,$M_\odot$ range;
    however, the range was 0.9-3.0\,$M_\odot$ in the present study, with only a
    few exceptions corresponding to $M>3.1\,M_\odot$.
These deviations can be interpreted by the facts that (1) our spectroscopic
    $T_{\rm eff}$ values obtained from low-resolution spectra were
    systematically lower than those reported in H2014, which were photometric
    $T_{\rm eff}$ from Sloan Digital Sky Survey {\it griz} filters \citep{
    Pinsonneault2012}; and (2) our approach correct the bias caused by
    different evolution speeds.
Therefore, the overall distribution shifted rightward towards the tracks of less
    massive evolved stars on the Kiel diagram.
Despite this, as $M\propto T_{\rm eff}^{3/2}$ whereas
    $\log{g}\propto0.5\log T_{\rm eff}$ according to Equations~\ref{para1} and
    \ref{para4}, and $\log{g}\propto\log{M/R^2}$ is explicitly related to
    $\nu_{\rm max}$ in Equation~\ref{numax}, independent of metallicity.
It is not surprising that a good agreement was found between the values of
    $\log{g}$ in the two studies.
    
Figures~\ref{fig_huber_cat01}-\ref{fig_huber_cat05} show that although
    asteroseismology yields satisfactory $\log{g}$ insensitive to $T_{\rm eff}$,
    the determination of stellar masses, and especially for giants, remains a
    challenge without reliable $T_{\rm eff}$ and [Fe/H].
Because the masses and radii of extra-solar planets are usually measured in
    terms of ratios relative to their host stars, the influence of inaccurate
    stellar $T_{\rm eff}$ and [Fe/H] is inevitable.
For giant stars, we estimate that an error of +100\,K in $T_{\rm eff}$ results
    in a mass error of about +0.20\,$M_\odot$ and a radius error of about
    +0.61\,$R_\odot$ when using the asteroseismic grid-based method.
In addition, an error of +0.1\,dex in [Fe/H] results in a mass error of
    +0.23\,$M_\odot$ and a radius error of about +0.74\,$R_\odot$.
This in turn emphasizes the importance of spectroscopic analysis of
    planet-hosting giant stars, for characterizing the planetary properties.
However, this effect is not significant for dwarfs.

\subsection{Calibration of the LAMOST $\log{g}$ Values}
\label{lamost_calib}

\begin{figure*}[htbp]
    \centering
    \includegraphics[width=12cm]{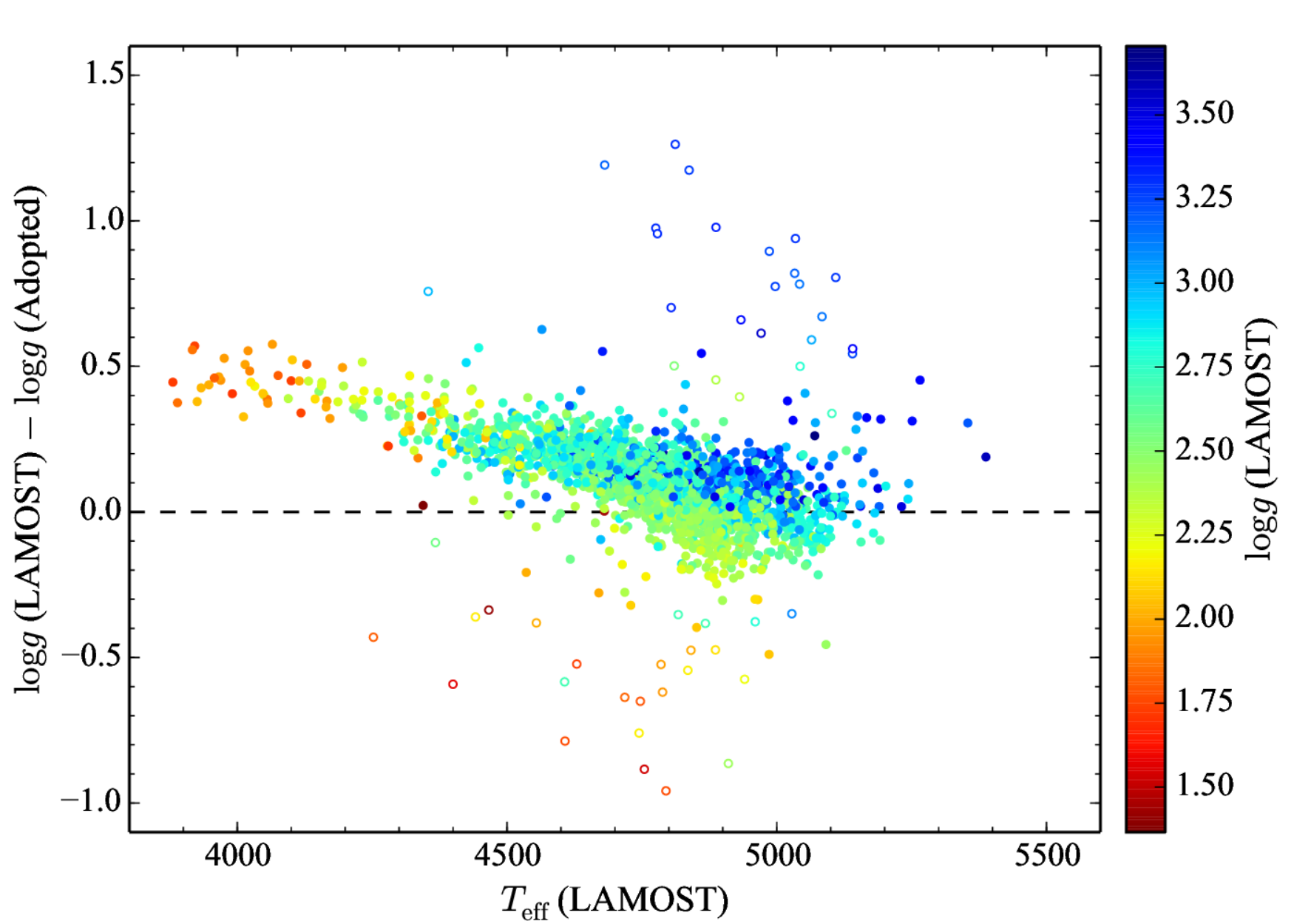}
    \caption{
        $\Delta\log{g}\,({\rm LAMOST - Adopted})$ vs. $T_{\rm eff}$ for 2,094
        giants.
        Color code correspond to the LAMOST $\log{g}$ values.
        Closed points are stars that were used in the least square fitting,
        while open circles represent the excluded outliers (see Text).
    }
    \label{dlogg_giants}
\end{figure*}

\begin{figure*}[htbp]
    \centering
    \includegraphics[width=12cm]{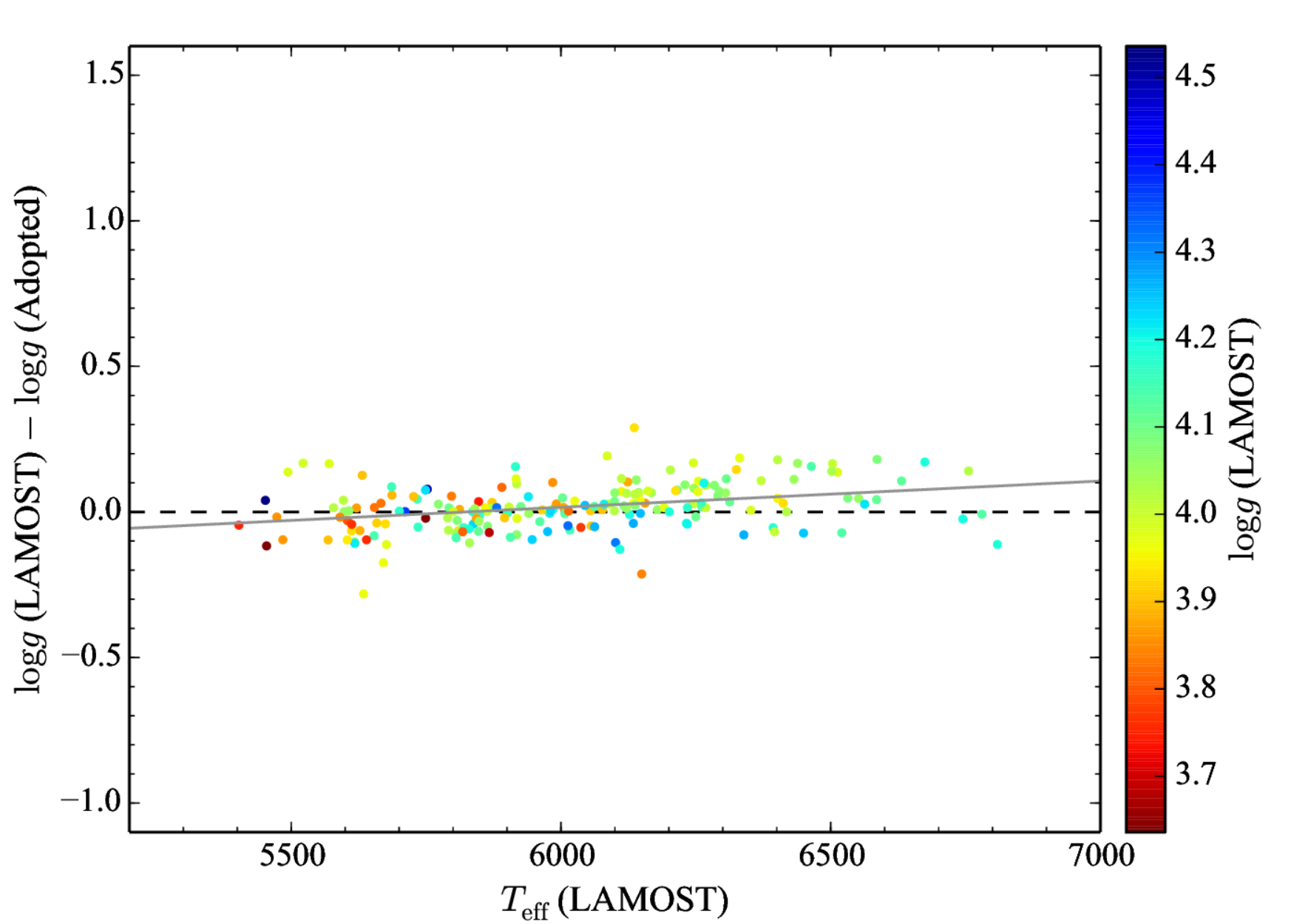}
    \caption{
        Similar to Figure~\ref{dlogg_giants} but for 195 dwarfs.
        The solid line is the best linear fit to the data points.
        The color code used in this figure is different from that in
        Figure~\ref{dlogg_giants}.
    }
    \label{dlogg_dwarfs}
\end{figure*}

By comparing the LAMOST $\log{g}$ values with asteroseismic $\log{g}$ values
    adopted in this work, we found that their difference exhibited a clear trend
    in the $T_{\rm eff}$ -- $\log{g}$ plane, implying a possibility to establish
    calibration relations for $\log{g}$ values of LAMOST samples.
To obtain reliable relationships, we excluded all spectra with SNR $<50$, and
    adopted the atmospheric parameters based on the spectra with highest SNR, if
    there were multiple observations for the same star in the LAMOST DR2 and
    DR3 Quarter 1 catalog.
This left us 2,289 samples, including 2,094 giants and 195 dwarfs.

\begin{figure*}
    \centering
    \includegraphics[height=5.4cm]{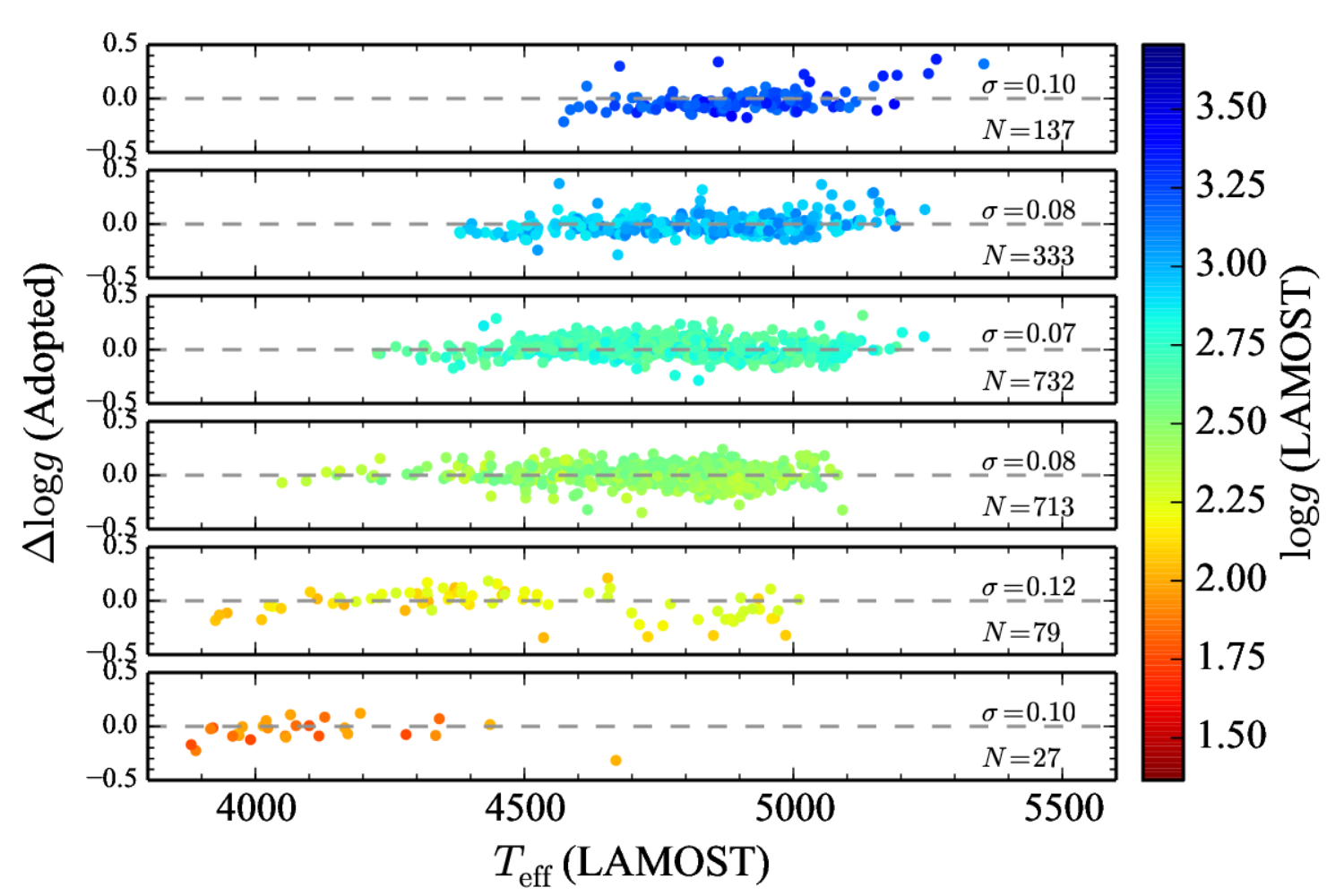}
    \includegraphics[height=5.4cm]{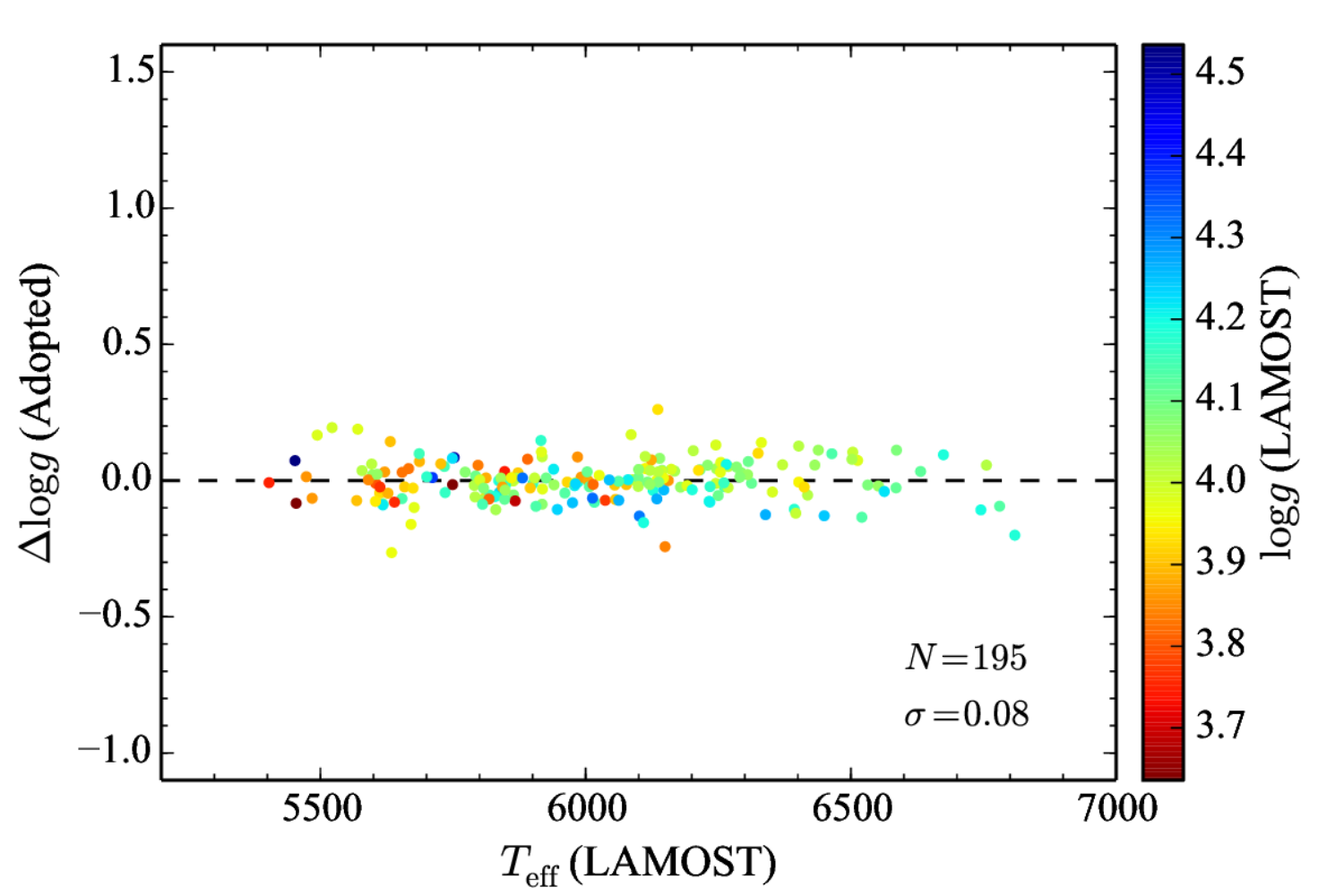}
    \caption{
        $\Delta\log{g}\,{\rm (Adopted)}$ for giants (left panel) and dwarfs
        (right panel) used in the least squares fitting.
        The residuals for giants are divided into six groups of $\log{g}$
        varying in steps of 0.3\,dex, from 1.7 (bottom panel) to 3.5 (top
        panel).
        The RMS values for data in each panel are also shown.
        }
    \label{fig_dlogg_res}
\end{figure*}

In Figure~\ref{dlogg_giants} we show the differences between $\log{g}\,_{\rm
    (LAMOST)}$ and $\log{g}\,_{\rm (Adopted)}$ as a function of $T_{\rm eff}$,
    with color coded by LAMOST $\log{g}$, for 2,094 giants with $T_{\rm eff}<$
    5,400\,K and $\log{g}<3.5$.
We used a first order 2D polynomial function $f(x,y)=p_0+p_1x+p_2y+p_3xy$ to
    model $\Delta\log{g}$, where $x$ is $T_{\rm eff}$ and $y$ is $\log{g}$.
The coefficients $p_0 \sim p_3$ were determined by least squares fitting.
After the coefficients were determined, the residuals of the fitting for all the
    data points were calculated.
In the next step, the points with residuals falling outside $\pm3\,\sigma$ were
    removed, and the least squares fitting was performed again.
The procedure converged after two iterations, when all residuals were within
    $\pm3\,\sigma$.
There are 2,044 stars left out of 2,094 giants.
This means that $\sim$2\% of the giants in Figure~\ref{dlogg_giants} are
    outliers that were not included in the fitting procedure.
The final relation was
\begin{multline}\label{cal_giants}
    \log{g}\,_{\rm (Adopted)} = \log{g}\,_{\rm (LAMOST)} \\
    - 5.716 + 1.283\times T_3 + 1.188\times\log{g}\,_{\rm (LAMOST)} \\
    - 0.2882\times T_3\times\log{g}\,_{\rm (LAMOST)}
\end{multline},
    where $T_3=T_{\rm eff} / 10^3$\,K is the normalized temperature from the
    LAMOST.
The range of temperatures in which this relation is applicable is 3,800\,K $\le
    T_{\rm eff}\le$ 4,500\,K for stars with $+1.3\le\log{g}\le2.2$, or 3,800\,K
    $\le T_{\rm eff}\le$ 5,200\,K for stars with $+2.2\le\log{g}\le3.5$.
In Figure~\ref{fig_dlogg_res}, we show the residuals of fitting, namely
    $\Delta\log{g}\,\rm{(Adopted)}$ for giants against $T_{\rm eff}\,{\rm
    (LAMOST)}$ within a range of 0.3\,dex in each panel, with $\log{g}$ in the
    1.7-3.5 range.
The RMS values of $\log{g}$ in each panel were comparable, varying from 0.07
    to 0.12\,dex.
We also calculated the RMS for these stars with $T_{\rm eff}$ in steps of
    200\,K, and obtained a 0.07-0.10\,dex variation range.
The overall RMS for all giants, excluding the outliers, was 0.082\,dex.

Figure~\ref{dlogg_dwarfs} shows the same relation for dwarfs with $T_{\rm eff}>$
    5,400\,K and $\log{g}>3.5$.
It is seen that $\Delta\log{g} ({\rm LAMOST} - {\rm Adopted})$ has a weak
    dependence on $T_{\rm eff}$, but no dependence on $\log{g}$.
Therefore, we only performed a linear least squares fit, which yielded
\begin{multline}\label{cal_dwarfs}
    \log{g}\,_{\rm (Adopted)} = \log{g}\,_{\rm (LAMOST)} \\
    +0.525 - 0.0902\times T_3
\end{multline},
where $T_3$ and $\log{g}\,_{\rm (LAMOST)}$ are the same as those in
    Equation~\ref{cal_giants}.
The applicable range is 5,400\,K $\le T_{\rm eff}\le$ 7,000\,K, and $+3.5\le
    \log{g}\le+4.5$.
The RMS value was only 0.075\,dex, and the residuals are plotted in
    Figure~\ref{fig_dlogg_res} (Right).

\begin{figure*}
    \centering
    \includegraphics[height=5.4cm]{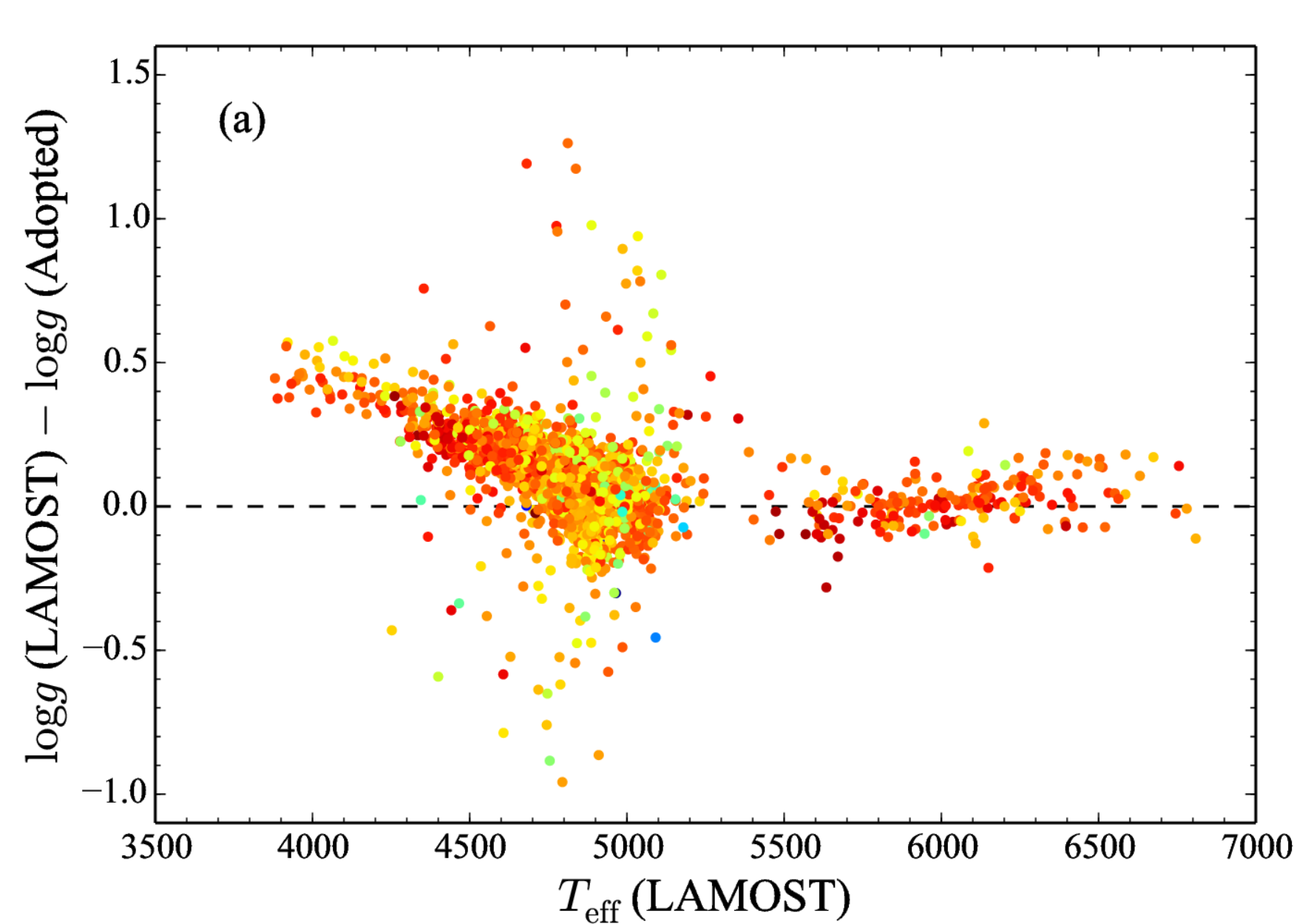}
    \includegraphics[height=5.4cm]{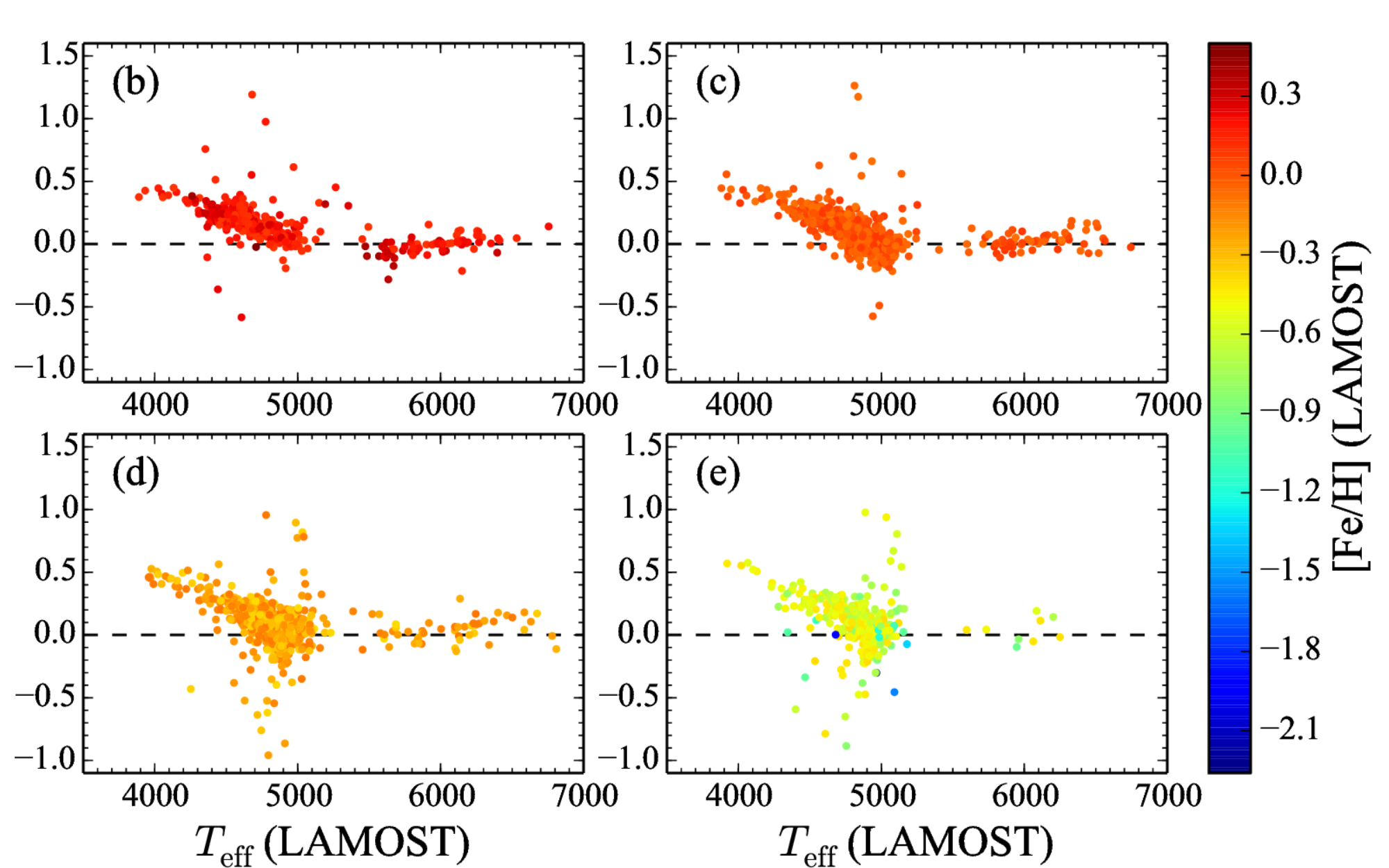}
    \caption{Panel~(a) shows the same relation as shown in
        Figures~\ref{dlogg_giants} and \ref{dlogg_dwarfs}, but with colors coded
        based on [Fe/H].
        The right panels show the sub-samples falling into four metallicity
        ranges:
        (b) [Fe/H] $>+0.1$;
        (c) $-0.1<$ [Fe/H] $<+0.1$;
        (d) $-0.4<$ [Fe/H] $<-0.1$; 
        and (e) [Fe/H] $<-0.4$.
        The number of stars in the four subsamples were 409, 702, 831, and 347,
        respectively.
        The right panels use the same color scale as panel~(a).
    }
    \label{fig_dlogg_feh}
\end{figure*}

In Figure~\ref{fig_dlogg_feh} we show the dependence of the relation on the
    stellar metallicity ([Fe/H]) as given in the LAMOST AFGK-type star
    parameters catalog.
We separated the entire sample into four groups, corresponding to different
    metallicity ranges:
    [Fe/H] $>+0.1$;
    $-0.1<$ [Fe/H] $<+0.1$;
    $-0.4<$ [Fe/H] $<-0.1$;
    and [Fe/H] $<-0.4$,
    as shown respectively in Panels~(b)-(e) of Figure~\ref{fig_dlogg_feh}.
Both the giants and dwarfs for different metallicities ranging from $-2.0$ to
    $+0.4$ exhibited very similar trends.
Although our entire sample covered a wide range of metallicity values, from
    $-2.26$ to $+0.50$, the number of stars at the metal-poor end was very small
    (only 10 stars with [Fe/H] $<-1.0$).
Therefore, the calibration relations are applicable for stars with $-1.0<$
    [Fe/H] $<+0.5$, and care must be taken for stars with metallicities outside
    this range.

\subsection{Calibration of the LAMOST $T_{\rm eff}$ and [Fe/H] Values}
\label{teff_calib}

\begin{figure*}[htbp]
    \centering
    \includegraphics[width=15cm]{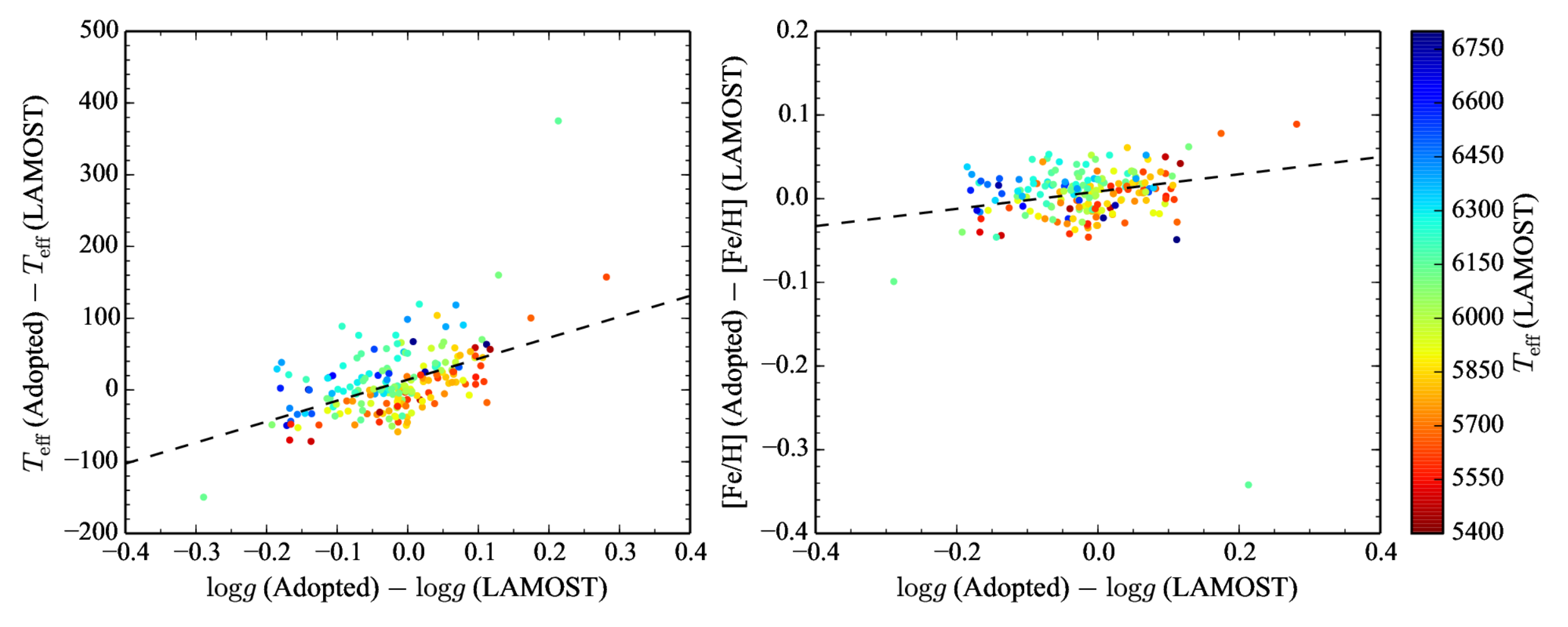}
    \caption{
        Same as Figure~\ref{dteff_giant} but for dwarfs.
        The dashed lines are the linear least squares fits.
    }
    \label{dteff_dwarf}
\end{figure*}

As shown in Section~\ref{lamost_calib}, the correction of LAMOST $\log{g}$
    reached 0.5\,dex for cool giants among our sample stars.
To quantify the impact of this change on $T_{\rm eff}$ and [Fe/H], we
    iteratively performed asteroseismic and spectroscopic analyses (see
    Section~\ref{iterative}).
Figures~\ref{dteff_giant} and \ref{dteff_dwarf} show the variations in
    $T_{\rm eff}$ and [Fe/H] vs. $\Delta\log{g}({\rm Adopted}-{\rm LAMOST})$
    for giants and dwarfs, respectively.
It is clear that the giants in Figure~\label{dteff_giant} can be divided into
    two groups.
Giants with $T_{\rm eff}>$ 4,500\,K exhibit clear correlations of both 
    $\Delta T_{\rm eff}$ and $\Delta$[Fe/H] with $\Delta\log{g}$.
Linear fits yielded
\begin{equation}\label{dteff_hot_giant}
    \Delta T_{\rm eff} = 284.9\times\Delta\log{g}\,_{\rm (Adopted - LAMOST)}
                         + 3.8
\end{equation}
and
\begin{equation}\label{dfeh_hot_giant}
    \Delta{\rm [Fe/H]} = 0.216\times\Delta\log{g}\,_{\rm (Adopted - LAMOST)}
                         + 0.008
\end{equation},
    where the finally adopted equations were $T_{\rm eff} =
    T_{\rm eff\,(LAMOST)} + \Delta T_{\rm eff}$ , and [Fe/H] =
    [Fe/H]$\,_{\rm (LAMOST)}$ + $\Delta$[Fe/H].
The RMS values for the above two equations were 18.4\,K and 0.026\,dex,
    respectively.
On the other hand, the $\Delta T_{\rm eff}$ for cooler giants ($T_{\rm eff}<$
    4,500\,K) exhibited no clear trends with
    $\Delta\log{g}({\rm Adopted}-{\rm LAMOST})$.

Figure~\ref{dteff_dwarf} shows that $\Delta T_{\rm eff}$ and $\Delta$[Fe/H] are
    significantly correlated with $\Delta\log{g}({\rm Adopted}-{\rm LAMOST})$
    for dwarfs.
Linear fits yielded
\begin{equation}\label{eqn_dteff_dwarf}
    \Delta T_{\rm eff} = 292.4\times\Delta\log{g}\,_{\rm (Adopted - LAMOST)} +
                         14.3
\end{equation}
    and
\begin{equation}\label{eqn_dfeh_dwarf}
    \Delta{\rm [Fe/H]} = 0.103\times\Delta\log{g}\,_{\rm (Adopted - LAMOST)} +
                         0.009
\end{equation},
    with the RMS values of 34.6\,K and 0.024\,dex, respectively.

\begin{figure*}[htbp]
    \centering
    \includegraphics[width=15cm]{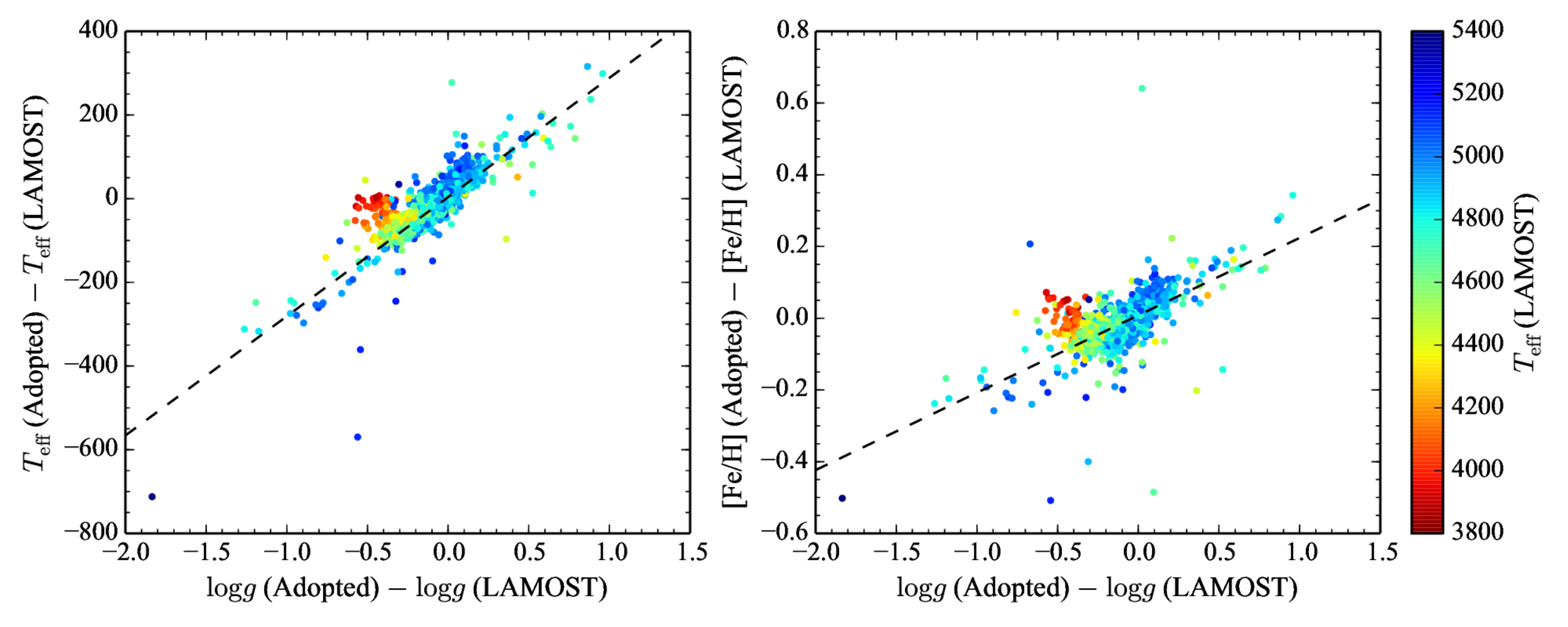}
    \caption{
        Variations in $T_{\rm eff}$ (left panel) and [Fe/H] (right panel) vs.
        $\Delta\log{g}$ for giants, with colors coded based on $T_{\rm eff}$.
        The dashed lines are the linear least squares fits for stars with
        $T_{\rm eff}>4,500$\,K (see Text).
    }
    \label{dteff_giant}
\end{figure*}

\subsection{Summary}
In this section, we propose empirical calibration relations for LAMOST $\log{g}$
    in Equations~\ref{cal_giants} and \ref{cal_dwarfs} for giants and dwarfs,
    respectively.
Generally speaking, the absolute values of $\log{g}$ corrections are much
    larger for giants than dwarfs, which reflects the difficulty associated with
    obtaining precise $\log{g}$ for evolved stars using low-resolution spectra.
For the coolest giants in our sample, with temperatures around 4,000\,K, the
    magnitude of corrections reached 0.5\,dex.
Because no systematic bias between asteroseismic $\log{g}$ and spectroscopic
    $\log{g}$ values has been found previously, the deviations of the LAMOST
    $\log{g}$ values from the asteroseismic ones are likely attributed to the
    adopted pipeline.
Because such major modifications of $\log{g}$ would inevitably affect the
    determination of both $T_{\rm eff}$ and [Fe/H], we provided the
    $T_{\rm eff}$ and [Fe/H] corrections in Equations~\ref{dteff_hot_giant} --
    \ref{eqn_dfeh_dwarf}.
Although the relations were derived for the {\em Kepler} targets, they are
    applicable to any LAMOST stars with spectroscopic parameters in the ranges
    given in Sections~\ref{lamost_calib} and \ref{teff_calib}.
The corrections of $\log{g}$, $T_{\rm eff}$ and [Fe/H] should always be applied
    together.

Due to their high luminosity, giant stars are visible from longer distances
    than dwarfs and play important roles in probing the Galactic structure,
    kinetics, and chemical evolution.
Because $g\sim MR^{-2}$, an overestimation of 0.5\,dex of $\log{g}$ for a K4
    giant with $T_{\rm eff}$ of 4,000\,K implies that the radius is
    underestimated by $0.5\ln(10)\Delta\log{g}\simeq$, or 58\% by assuming a
    fixed $M$.
This in turn causes $\sim$115\% underestimation of stellar luminosity as
    $L\sim R^2T_{\rm eff}^4$.
Here $T_{\rm eff}$ is fixed because, according to Figure~\ref{dteff_giant}, it
    nearly does not change with $\log{g}$.
Furthermore, considering that $L=4\pi D^2F$ , where $D$ denotes the distance and
    $F$ is the observed flux density, the luminosity distance is also
    underestimated by $\sim$58\% if the interstellar extinction is ignored.
For a typical K1 giant with $T_{\rm eff}$ of 4,600\,K and $\log{g}$ of 2.8,
    LAMOST overestimates its actual $\log{g}$ by $\sim$0.22\,dex; consequently,
    the values of $R$, $L$, and $D$ will be underestimated by 25\%, 50\%, and
    25\%, respectively.
However, given that $T_{\rm eff}$ is also reduced by $\sim$56\,K (according to
    Equation~\ref{dteff_giant}), the impact of increasing $R$ on $L$ will be
    offset by $\sim$5\%.
Therefore, the resulting luminosity and distance need to be increased by 45\%
    and 22\%, respectively.
Another example is a red-clump giant with $T_{\rm eff}=4,900$ K and $\log{g}=$
    2.6, for which the correction of $\log{g}$ is close to zero.
Thus, previous works based on LAMOST red-clump giants \citep[e.g.][]{Wan2015}
    are nearly not affected by the systematic deviations of $\log{g}$.

\section{CONCLUSIONS}
In this paper, we consistently derived stellar parameters for a large sample of
    stars with the oscillation data from the {\em Kepler} mission, along with
    the $T_{\rm eff}$ and [Fe/H] values from the LAMOST low-resolution
    ($R\sim1,800$) spectra.
Spanning a wide range of metallicity values ($-2.3<$ [Fe/H] $<+0.5$), the entire
    sample contained 2,831 giants and 229 dwarfs, of which 15 have been
    confirmed to harbor extra-solar planets and 22 were potential planet-host
    candidates. 
The stellar properties were calculated using an improved grid-based method, by
    considering the evolution speed effect and the post-RGB phases.
The fact that $T_{\rm eff}$ and [Fe/H] values were derived from the spectra with
    SNR $>$ 30 and $\log{g}$ values were derived from the {\em Kepler}
    oscillation parameters ensure the accuracy of our results, compensating for
    the shortage of low-resolution spectroscopy.
By comparing the asteroseismology and spectroscopic results, we found that
    LAMOST yielded systematically higher $\log{g}$ for giants, and the
    overestimation exhibited clear trends with $T_{\rm eff}$ and $\log{g}$.
We established calibration relations for the $\log{g}$ of LAMOST，for both
    giants and dwarfs.
The post-calibration uncertainty in $\log{g}$ was 0.08\,dex for both giants and
    dwarfs, corresponding to distance errors of only 8\%.
The empirical relations were established for a range of stars, from mildly
    metal-poor ([Fe/H] $\sim -1.0$) to those with super-solar metallicity
    ([Fe/H] $\sim+0.4$).
This range covers most of the giants and FGK dwarfs that have been observed by
    LAMOST.
We suggest that $\log{g}$ of stars in this metallicity range should be corrected
    by using our derived relations.
Meanwhile, our results regarding stellar physical parameters show that
    photometric $T_{\rm eff}$ and [Fe/H] are not sufficiently accurate for
    obtaining reliable masses and radii for giants, even when augmented by
    global asteroseismic quantities.
Therefore, spectroscopic studies are critical for characterization of these
    parameters.
    
\acknowledgements
This research is supported by the National Natural Science Foundation of China
    under grants No. 11390371 and 11403056.
The Guoshoujing Telescope (the Large Sky Area Multi-Object Fiber Spectroscopic
    Telescope, LAMOST) is a National Major Scientific Project built by the
    Chinese Academy of Sciences.
Funding for the project has been provided by the National Development and Reform
    Commission.
LAMOST is operated and managed by the National Astronomical Observatories,
    Chinese Academy of Sciences. 
LW thanks Tim Bedding, William Chaplin, Saskia Hekker, Daniel Huber, Kalinger
    and Dennis Stello for providing the seismic data, and Tanda Li for
    helpful discussions.

\bibliography{references}

\begin{thebibliography}{}
\expandafter\ifx\csname natexlab\endcsname\relax\def\natexlab#1{#1}\fi

\bibitem[{{Alexander} \& {Ferguson}(1994)}]{Alexander1994}
{Alexander}, D.~R., \& {Ferguson}, J.~W. 1994, \apj, 437, 879

\bibitem[{{Angus} {et~al.}(2015){Angus}, {Aigrain}, {Foreman-Mackey}, \&
  {McQuillan}}]{Angus2015}
{Angus}, R., {Aigrain}, S., {Foreman-Mackey}, D., \& {McQuillan}, A. 2015,
  \mnras, 450, 1787

\bibitem[{{Basu} {et~al.}(2010){Basu}, {Chaplin}, \& {Elsworth}}]{Basu2010}
{Basu}, S., {Chaplin}, W.~J., \& {Elsworth}, Y. 2010, \apj, 710, 1596

\bibitem[{{Boesgaard} {et~al.}(2011){Boesgaard}, {Rich}, {Levesque}, \&
  {Bowler}}]{Boesgaard2011}
{Boesgaard}, A.~M., {Rich}, J.~A., {Levesque}, E.~M., \& {Bowler}, B.~P. 2011,
  \apj, 743, 140

\bibitem[{{Borucki} {et~al.}(2010){Borucki}, {Koch}, {Basri}, {Batalha},
  {Brown}, {Caldwell}, {Caldwell}, {Christensen-Dalsgaard}, {Cochran},
  {DeVore}, {Dunham}, {Dupree}, {Gautier}, {Geary}, {Gilliland}, {Gould},
  {Howell}, {Jenkins}, {Kondo}, {Latham}, {Marcy}, {Meibom}, {Kjeldsen},
  {Lissauer}, {Monet}, {Morrison}, {Sasselov}, {Tarter}, {Boss}, {Brownlee},
  {Owen}, {Buzasi}, {Charbonneau}, {Doyle}, {Fortney}, {Ford}, {Holman},
  {Seager}, {Steffen}, {Welsh}, {Rowe}, {Anderson}, {Buchhave}, {Ciardi},
  {Walkowicz}, {Sherry}, {Horch}, {Isaacson}, {Everett}, {Fischer}, {Torres},
  {Johnson}, {Endl}, {MacQueen}, {Bryson}, {Dotson}, {Haas}, {Kolodziejczak},
  {Van Cleve}, {Chandrasekaran}, {Twicken}, {Quintana}, {Clarke}, {Allen},
  {Li}, {Wu}, {Tenenbaum}, {Verner}, {Bruhweiler}, {Barnes}, \&
  {Prsa}}]{Borucki2010}
{Borucki}, W.~J., {Koch}, D., {Basri}, G., {et~al.} 2010, Science, 327, 977

\bibitem[{{Breddels} {et~al.}(2010){Breddels}, {Smith}, {Helmi},
  {Bienaym{\'e}}, {Binney}, {Bland-Hawthorn}, {Boeche}, {Burnett}, {Campbell},
  {Freeman}, {Gibson}, {Gilmore}, {Grebel}, {Munari}, {Navarro}, {Parker},
  {Seabroke}, {Siebert}, {Siviero}, {Steinmetz}, {Watson}, {Williams}, {Wyse},
  \& {Zwitter}}]{Breddels2010}
{Breddels}, M.~A., {Smith}, M.~C., {Helmi}, A., {et~al.} 2010, \aap, 511, A90

\bibitem[{{Brown} {et~al.}(1991){Brown}, {Gilliland}, {Noyes}, \&
  {Ramsey}}]{Brown1991}
{Brown}, T.~M., {Gilliland}, R.~L., {Noyes}, R.~W., \& {Ramsey}, L.~W. 1991,
  \apj, 368, 599

\bibitem[{{Brown} {et~al.}(2011){Brown}, {Latham}, {Everett}, \&
  {Esquerdo}}]{Brown2011}
{Brown}, T.~M., {Latham}, D.~W., {Everett}, M.~E., \& {Esquerdo}, G.~A. 2011,
  \aj, 142, 112

\bibitem[{{Bruntt} {et~al.}(2012){Bruntt}, {Basu}, {Smalley}, {Chaplin},
  {Verner}, {Bedding}, {Catala}, {Gazzano}, {Molenda-{\.Z}akowicz}, {Thygesen},
  {Uytterhoeven}, {Hekker}, {Huber}, {Karoff}, {Mathur}, {Mosser},
  {Appourchaux}, {Campante}, {Elsworth}, {Garc{\'{\i}}a}, {Handberg},
  {Metcalfe}, {Quirion}, {R{\'e}gulo}, {Roxburgh}, {Stello},
  {Christensen-Dalsgaard}, {Kawaler}, {Kjeldsen}, {Morris}, {Quintana}, \&
  {Sanderfer}}]{Bruntt2012}
{Bruntt}, H., {Basu}, S., {Smalley}, B., {et~al.} 2012, \mnras, 423, 122

\bibitem[{{Buchhave} {et~al.}(2012){Buchhave}, {Latham}, {Johansen},
  {Bizzarro}, {Torres}, {Rowe}, {Batalha}, {Borucki}, {Brugamyer}, {Caldwell},
  {Bryson}, {Ciardi}, {Cochran}, {Endl}, {Esquerdo}, {Ford}, {Geary},
  {Gilliland}, {Hansen}, {Isaacson}, {Laird}, {Lucas}, {Marcy}, {Morse},
  {Robertson}, {Shporer}, {Stefanik}, {Still}, \& {Quinn}}]{Buchhave2012}
{Buchhave}, L.~A., {Latham}, D.~W., {Johansen}, A., {et~al.} 2012, \nat, 486,
  375

\bibitem[{{Carlin} {et~al.}(2015){Carlin}, {Liu}, {Newberg}, {Beers}, {Chen},
  {Deng}, {Guhathakurta}, {Hou}, {Hou}, {L{\'e}pine}, {Li}, {Luo}, {Smith},
  {Wu}, {Yang}, {Yanny}, {Zhang}, \& {Zheng}}]{Carlin2015}
{Carlin}, J.~L., {Liu}, C., {Newberg}, H.~J., {et~al.} 2015, \aj, 150, 4

\bibitem[{{Cenarro} {et~al.}(2007){Cenarro}, {Peletier},
  {S{\'a}nchez-Bl{\'a}zquez}, {Selam}, {Toloba}, {Cardiel},
  {Falc{\'o}n-Barroso}, {Gorgas}, {Jim{\'e}nez-Vicente}, \&
  {Vazdekis}}]{Cenarro2007}
{Cenarro}, A.~J., {Peletier}, R.~F., {S{\'a}nchez-Bl{\'a}zquez}, P., {et~al.}
  2007, \mnras, 374, 664

\bibitem[{{Chaplin} {et~al.}(2011){Chaplin}, {Kjeldsen},
  {Christensen-Dalsgaard}, {Basu}, {Miglio}, {Appourchaux}, {Bedding},
  {Elsworth}, {Garc{\'{\i}}a}, {Gilliland}, {Girardi}, {Houdek}, {Karoff},
  {Kawaler}, {Metcalfe}, {Molenda-{\.Z}akowicz}, {Monteiro}, {Thompson},
  {Verner}, {Ballot}, {Bonanno}, {Brand{\~a}o}, {Broomhall}, {Bruntt},
  {Campante}, {Corsaro}, {Creevey}, {Do{\u g}an}, {Esch}, {Gai}, {Gaulme},
  {Hale}, {Handberg}, {Hekker}, {Huber}, {Jim{\'e}nez}, {Mathur}, {Mazumdar},
  {Mosser}, {New}, {Pinsonneault}, {Pricopi}, {Quirion}, {R{\'e}gulo},
  {Salabert}, {Serenelli}, {Silva Aguirre}, {Sousa}, {Stello}, {Stevens},
  {Suran}, {Uytterhoeven}, {White}, {Borucki}, {Brown}, {Jenkins}, {Kinemuchi},
  {Van Cleve}, \& {Klaus}}]{Chaplin2011}
{Chaplin}, W.~J., {Kjeldsen}, H., {Christensen-Dalsgaard}, J., {et~al.} 2011,
  Science, 332, 213

\bibitem[{{Chaplin} {et~al.}(2013){Chaplin}, {Sanchis-Ojeda}, {Campante},
  {Handberg}, {Stello}, {Winn}, {Basu}, {Christensen-Dalsgaard}, {Davies},
  {Metcalfe}, {Buchhave}, {Fischer}, {Bedding}, {Cochran}, {Elsworth},
  {Gilliland}, {Hekker}, {Huber}, {Isaacson}, {Karoff}, {Kawaler}, {Kjeldsen},
  {Latham}, {Lund}, {Lundkvist}, {Marcy}, {Miglio}, {Barclay}, \&
  {Lissauer}}]{Chaplin2013}
{Chaplin}, W.~J., {Sanchis-Ojeda}, R., {Campante}, T.~L., {et~al.} 2013, \apj,
  766, 101

\bibitem[{{Chaplin} {et~al.}(2014){Chaplin}, {Basu}, {Huber}, {Serenelli},
  {Casagrande}, {Silva Aguirre}, {Ball}, {Creevey}, {Gizon}, {Handberg},
  {Karoff}, {Lutz}, {Marques}, {Miglio}, {Stello}, {Suran}, {Pricopi},
  {Metcalfe}, {Monteiro}, {Molenda-{\.Z}akowicz}, {Appourchaux},
  {Christensen-Dalsgaard}, {Elsworth}, {Garc{\'{\i}}a}, {Houdek}, {Kjeldsen},
  {Bonanno}, {Campante}, {Corsaro}, {Gaulme}, {Hekker}, {Mathur}, {Mosser},
  {R{\'e}gulo}, \& {Salabert}}]{Chaplin2014}
{Chaplin}, W.~J., {Basu}, S., {Huber}, D., {et~al.} 2014, \apjs, 210, 1

\bibitem[{{Chen} {et~al.}(2000){Chen}, {Nissen}, {Zhao}, {Zhang}, \&
  {Benoni}}]{Chen2000}
{Chen}, Y.~Q., {Nissen}, P.~E., {Zhao}, G., {Zhang}, H.~W., \& {Benoni}, T.
  2000, \aaps, 141, 491

\bibitem[{{Chmielewski}(2000)}]{Chmielewski2000}
{Chmielewski}, Y. 2000, \aap, 353, 666

\bibitem[{{Cui} {et~al.}(2012){Cui}, {Zhao}, {Chu}, {Li}, {Li}, {Zhang}, {Su},
  {Yao}, {Wang}, {Xing}, {Li}, {Zhu}, {Wang}, {Gu}, {Luo}, {Xu}, {Zhang},
  {Liu}, {Zhang}, {Yang}, {Cao}, {Chen}, {Chen}, {Chen}, {Chen}, {Chu}, {Feng},
  {Gong}, {Hou}, {Hu}, {Hu}, {Hu}, {Jia}, {Jiang}, {Jiang}, {Jiang}, {Jin},
  {Li}, {Li}, {Li}, {Liu}, {Liu}, {Lu}, {Mao}, {Men}, {Qi}, {Qi}, {Shi},
  {Tang}, {Tao}, {Wang}, {Wang}, {Wang}, {Wang}, {Wang}, {Wang}, {Wang},
  {Wang}, {Wang}, {Wang}, {Wang}, {Wang}, {Xu}, {Xu}, {Yang}, {Yu}, {Yuan},
  {Yuan}, {Zhai}, {Zhang}, {Zhang}, {Zhang}, {Zhao}, {Zhou}, {Zhou}, {Zhu}, \&
  {Zou}}]{Cui2012}
{Cui}, X.-Q., {Zhao}, Y.-H., {Chu}, Y.-Q., {et~al.} 2012, Research in Astronomy
  and Astrophysics, 12, 1197

\bibitem[{{da Silva} {et~al.}(2006){da Silva}, {Girardi}, {Pasquini},
  {Setiawan}, {von der L{\"u}he}, {de Medeiros}, {Hatzes}, {D{\"o}llinger}, \&
  {Weiss}}]{daSilva2006}
{da Silva}, L., {Girardi}, L., {Pasquini}, L., {et~al.} 2006, \aap, 458, 609

\bibitem[{{De Cat} {et~al.}(2015){De Cat}, {Fu}, {Ren}, {Yang}, {Shi}, {Luo},
  {Yang}, {Wang}, {Zhang}, {Shi}, {Zhang}, {Dong}, {Catanzaro}, {Corbally},
  {Frasca}, {Gray}, {Molenda-{\.Z}akowicz}, {Uytterhoeven}, {Briquet},
  {Bruntt}, {Frandsen}, {Kiss}, {Kurtz}, {Marconi}, {Niemczura}, {{\O}stensen},
  {Ripepi}, {Smalley}, {Southworth}, {Szab{\'o}}, {Telting}, {Karoff}, {Silva
  Aguirre}, {Wu}, {Hou}, {Jin}, \& {Zhou}}]{DeCat2015}
{De Cat}, P., {Fu}, J.~N., {Ren}, A.~B., {et~al.} 2015, \apjs, 220, 19

\bibitem[{{de Jager} {et~al.}(1988){de Jager}, {Nieuwenhuijzen}, \& {van der
  Hucht}}]{Jager1988}
{de Jager}, C., {Nieuwenhuijzen}, H., \& {van der Hucht}, K.~A. 1988, \aaps,
  72, 259

\bibitem[{{Deeming}(1960)}]{Deeming1960}
{Deeming}, T.~J. 1960, \mnras, 121, 52

\bibitem[{{Demarque} {et~al.}(2008){Demarque}, {Guenther}, {Li}, {Mazumdar}, \&
  {Straka}}]{Demarque2008}
{Demarque}, P., {Guenther}, D.~B., {Li}, L.~H., {Mazumdar}, A., \& {Straka},
  C.~W. 2008, \apss, 316, 31

\bibitem[{{Dong} {et~al.}(2014){Dong}, {Zheng}, {Zhu}, {De Cat}, {Fu}, {Yang},
  {Zhang}, {Jin}, \& {Zhang}}]{Dong2014}
{Dong}, S., {Zheng}, Z., {Zhu}, Z., {et~al.} 2014, \apjl, 789, L3

\bibitem[{{Dotter} {et~al.}(2008){Dotter}, {Chaboyer}, {Jevremovi{\'c}},
  {Kostov}, {Baron}, \& {Ferguson}}]{Dotter2008}
{Dotter}, A., {Chaboyer}, B., {Jevremovi{\'c}}, D., {et~al.} 2008, \apjs, 178,
  89

\bibitem[{{Fuhrmann}(1998)}]{Fuhrmann1998}
{Fuhrmann}, K. 1998, \aap, 338, 161

\bibitem[{{Gai} {et~al.}(2011){Gai}, {Basu}, {Chaplin}, \&
  {Elsworth}}]{Gai2011}
{Gai}, N., {Basu}, S., {Chaplin}, W.~J., \& {Elsworth}, Y. 2011, \apj, 730, 63

\bibitem[{{Garc{\'{\i}}a} {et~al.}(2014){Garc{\'{\i}}a}, {Ceillier},
  {Salabert}, {Mathur}, {van Saders}, {Pinsonneault}, {Ballot}, {Beck},
  {Bloemen}, {Campante}, {Davies}, {do Nascimento}, {Mathis}, {Metcalfe},
  {Nielsen}, {Su{\'a}rez}, {Chaplin}, {Jim{\'e}nez}, \& {Karoff}}]{Garcia2014}
{Garc{\'{\i}}a}, R.~A., {Ceillier}, T., {Salabert}, D., {et~al.} 2014, \aap,
  572, A34

\bibitem[{{Hekker} {et~al.}(2013){Hekker}, {Elsworth}, {Mosser}, {Kallinger},
  {Basu}, {Chaplin}, \& {Stello}}]{Hekker2013}
{Hekker}, S., {Elsworth}, Y., {Mosser}, B., {et~al.} 2013, \aap, 556, A59

\bibitem[{{Hekker} {et~al.}(2010){Hekker}, {Broomhall}, {Chaplin}, {Elsworth},
  {Fletcher}, {New}, {Arentoft}, {Quirion}, \& {Kjeldsen}}]{Hekker2010}
{Hekker}, S., {Broomhall}, A.-M., {Chaplin}, W.~J., {et~al.} 2010, \mnras, 402,
  2049

\bibitem[{{Hekker} {et~al.}(2011){Hekker}, {Gilliland}, {Elsworth}, {Chaplin},
  {De Ridder}, {Stello}, {Kallinger}, {Ibrahim}, {Klaus}, \& {Li}}]{Hekker2011}
{Hekker}, S., {Gilliland}, R.~L., {Elsworth}, Y., {et~al.} 2011, \mnras, 414,
  2594

\bibitem[{{Hirano} {et~al.}(2012){Hirano}, {Sanchis-Ojeda}, {Takeda}, {Narita},
  {Winn}, {Taruya}, \& {Suto}}]{Hirano2012}
{Hirano}, T., {Sanchis-Ojeda}, R., {Takeda}, Y., {et~al.} 2012, \apj, 756, 66

\bibitem[{{Hirano} {et~al.}(2014){Hirano}, {Sanchis-Ojeda}, {Takeda}, {Winn},
  {Narita}, \& {Takahashi}}]{Hirano2014}
---. 2014, \apj, 783, 9

\bibitem[{{Huber} {et~al.}(2009){Huber}, {Stello}, {Bedding}, {Chaplin},
  {Arentoft}, {Quirion}, \& {Kjeldsen}}]{Huber2009}
{Huber}, D., {Stello}, D., {Bedding}, T.~R., {et~al.} 2009, Communications in
  Asteroseismology, 160, 74

\bibitem[{{Huber} {et~al.}(2011){Huber}, {Bedding}, {Stello}, {Hekker},
  {Mathur}, {Mosser}, {Verner}, {Bonanno}, {Buzasi}, {Campante}, {Elsworth},
  {Hale}, {Kallinger}, {Silva Aguirre}, {Chaplin}, {De Ridder},
  {Garc{\'{\i}}a}, {Appourchaux}, {Frandsen}, {Houdek}, {Molenda-{\.Z}akowicz},
  {Monteiro}, {Christensen-Dalsgaard}, {Gilliland}, {Kawaler}, {Kjeldsen},
  {Broomhall}, {Corsaro}, {Salabert}, {Sanderfer}, {Seader}, \&
  {Smith}}]{Huber2011}
{Huber}, D., {Bedding}, T.~R., {Stello}, D., {et~al.} 2011, \apj, 743, 143

\bibitem[{{Huber} {et~al.}(2013){Huber}, {Chaplin}, {Christensen-Dalsgaard},
  {Gilliland}, {Kjeldsen}, {Buchhave}, {Fischer}, {Lissauer}, {Rowe},
  {Sanchis-Ojeda}, {Basu}, {Handberg}, {Hekker}, {Howard}, {Isaacson},
  {Karoff}, {Latham}, {Lund}, {Lundkvist}, {Marcy}, {Miglio}, {Silva Aguirre},
  {Stello}, {Arentoft}, {Barclay}, {Bedding}, {Burke}, {Christiansen},
  {Elsworth}, {Haas}, {Kawaler}, {Metcalfe}, {Mullally}, \&
  {Thompson}}]{Huber2013}
{Huber}, D., {Chaplin}, W.~J., {Christensen-Dalsgaard}, J., {et~al.} 2013,
  \apj, 767, 127

\bibitem[{{Huber} {et~al.}(2014){Huber}, {Silva Aguirre}, {Matthews},
  {Pinsonneault}, {Gaidos}, {Garc{\'{\i}}a}, {Hekker}, {Mathur}, {Mosser},
  {Torres}, {Bastien}, {Basu}, {Bedding}, {Chaplin}, {Demory}, {Fleming},
  {Guo}, {Mann}, {Rowe}, {Serenelli}, {Smith}, \& {Stello}}]{Huber2014}
{Huber}, D., {Silva Aguirre}, V., {Matthews}, J.~M., {et~al.} 2014, \apjs, 211,
  2

\bibitem[{{Iglesias} \& {Rogers}(1996)}]{Iglesias1996}
{Iglesias}, C.~A., \& {Rogers}, F.~J. 1996, \apj, 464, 943

\bibitem[{{Johnson} {et~al.}(2014){Johnson}, {Huber}, {Boyajian}, {Brewer},
  {White}, {von Braun}, {Maestro}, {Stello}, \& {Barclay}}]{Johnson2014}
{Johnson}, J.~A., {Huber}, D., {Boyajian}, T., {et~al.} 2014, \apj, 794, 15

\bibitem[{{Kallinger} {et~al.}(2010{\natexlab{a}}){Kallinger}, {Mosser},
  {Hekker}, {Huber}, {Stello}, {Mathur}, {Basu}, {Bedding}, {Chaplin}, {De
  Ridder}, {Elsworth}, {Frandsen}, {Garc{\'{\i}}a}, {Gruberbauer}, {Matthews},
  {Borucki}, {Bruntt}, {Christensen-Dalsgaard}, {Gilliland}, {Kjeldsen}, \&
  {Koch}}]{Kallinger2010}
{Kallinger}, T., {Mosser}, B., {Hekker}, S., {et~al.} 2010{\natexlab{a}}, \aap,
  522, A1

\bibitem[{{Kallinger} {et~al.}(2010{\natexlab{b}}){Kallinger}, {Weiss},
  {Barban}, {Baudin}, {Cameron}, {Carrier}, {De Ridder}, {Goupil},
  {Gruberbauer}, {Hatzes}, {Hekker}, {Samadi}, \& {Deleuil}}]{Kallinger2010a}
{Kallinger}, T., {Weiss}, W.~W., {Barban}, C., {et~al.} 2010{\natexlab{b}},
  \aap, 509, A77

\bibitem[{{Kjeldsen} \& {Bedding}(1995)}]{Kjeldsen1995}
{Kjeldsen}, H., \& {Bedding}, T.~R. 1995, \aap, 293, 87

\bibitem[{{Kurucz}(1991)}]{Kurucz1991}
{Kurucz}, R.~L. 1991, in NATO Advanced Science Institutes (ASI) Series C, Vol.
  341, NATO Advanced Science Institutes (ASI) Series C, ed. L.~{Crivellari},
  I.~{Hubeny}, \& D.~G. {Hummer}, 441

\bibitem[{{Lee} {et~al.}(2008){Lee}, {Beers}, {Sivarani}, {Allende Prieto},
  {Koesterke}, {Wilhelm}, {Re Fiorentin}, {Bailer-Jones}, {Norris}, {Rockosi},
  {Yanny}, {Newberg}, {Covey}, {Zhang}, \& {Luo}}]{Lee2008}
{Lee}, Y.~S., {Beers}, T.~C., {Sivarani}, T., {et~al.} 2008, \aj, 136, 2022

\bibitem[{{Lejeune} \& {Schaerer}(2001)}]{Lejeune2001}
{Lejeune}, T., \& {Schaerer}, D. 2001, \aap, 366, 538

\bibitem[{{Lind} {et~al.}(2012){Lind}, {Bergemann}, \& {Asplund}}]{Lind2012}
{Lind}, K., {Bergemann}, M., \& {Asplund}, M. 2012, \mnras, 427, 50

\bibitem[{{Liu} {et~al.}(2015){Liu}, {Fang}, {Wu}, {Deng}, {Wang}, {Wang},
  {Fu}, {Hou}, {Li}, \& {Zhang}}]{Liu2015}
{Liu}, C., {Fang}, M., {Wu}, Y., {et~al.} 2015, \apj, 807, 4

\bibitem[{{Luo} {et~al.}(2015){Luo}, {Zhao}, {Zhao}, {Deng}, {Liu}, {Jing},
  {Wang}, {Zhang}, {Shi}, {Cui}, {Chu}, {Li}, {Bai}, {Wu}, {Cai}, {Cao}, {Cao},
  {Carlin}, {Chen}, {Chen}, {Chen}, {Chen}, {Chen}, {Chen}, {Chen},
  {Christlieb}, {Chu}, {Cui}, {Dong}, {Du}, {Fan}, {Feng}, {Fu}, {Gao}, {Gong},
  {Gu}, {Guo}, {Han}, {He}, {Hou}, {Hou}, {Hou}, {Hu}, {Hu}, {Hu}, {Huo},
  {Jia}, {Jiang}, {Jiang}, {Jiang}, {Jin}, {Kong}, {Kong}, {Lei}, {Li}, {Li},
  {Li}, {Li}, {Li}, {Li}, {Li}, {Li}, {Li}, {Li}, {Li}, {Li}, {Liang}, {Lin},
  {Liu}, {Liu}, {Liu}, {Liu}, {Lu}, {Luo}, {Mao}, {Newberg}, {Ni}, {Qi}, {Qi},
  {Shen}, {Shi}, {Song}, {Song}, {Su}, {Su}, {Tang}, {Tao}, {Tian}, {Wang},
  {Wang}, {Wang}, {Wang}, {Wang}, {Wang}, {Wang}, {Wang}, {Wang}, {Wang},
  {Wang}, {Wang}, {Wang}, {Wang}, {Wang}, {Wang}, {Wang}, {Wang}, {Wang},
  {Wang}, {Wei}, {Wei}, {Wu}, {Wu}, {Wu}, {Wu}, {Xing}, {Xu}, {Xu}, {Xu},
  {Yan}, {Yang}, {Yang}, {Yang}, {Yang}, {Yao}, {Yu}, {Yuan}, {Yuan}, {Yuan},
  {Yuan}, {Zhai}, {Zhang}, {Zhang}, {Zhang}, {Zhang}, {Zhang}, {Zhang},
  {Zhang}, {Zhang}, {Zhao}, {Zhou}, {Zhou}, {Zhu}, {Zhu}, {Zou}, \&
  {Zuo}}]{Luo2015}
{Luo}, A.-L., {Zhao}, Y.-H., {Zhao}, G., {et~al.} 2015, Research in Astronomy
  and Astrophysics, 15, 1095

\bibitem[{{Marcy} {et~al.}(2014){Marcy}, {Isaacson}, {Howard}, {Rowe},
  {Jenkins}, {Bryson}, {Latham}, {Howell}, {Gautier}, {Batalha}, {Rogers},
  {Ciardi}, {Fischer}, {Gilliland}, {Kjeldsen}, {Christensen-Dalsgaard},
  {Huber}, {Chaplin}, {Basu}, {Buchhave}, {Quinn}, {Borucki}, {Koch}, {Hunter},
  {Caldwell}, {Van Cleve}, {Kolbl}, {Weiss}, {Petigura}, {Seager}, {Morton},
  {Johnson}, {Ballard}, {Burke}, {Cochran}, {Endl}, {MacQueen}, {Everett},
  {Lissauer}, {Ford}, {Torres}, {Fressin}, {Brown}, {Steffen}, {Charbonneau},
  {Basri}, {Sasselov}, {Winn}, {Sanchis-Ojeda}, {Christiansen}, {Adams},
  {Henze}, {Dupree}, {Fabrycky}, {Fortney}, {Tarter}, {Holman}, {Tenenbaum},
  {Shporer}, {Lucas}, {Welsh}, {Orosz}, {Bedding}, {Campante}, {Davies},
  {Elsworth}, {Handberg}, {Hekker}, {Karoff}, {Kawaler}, {Lund}, {Lundkvist},
  {Metcalfe}, {Miglio}, {Silva Aguirre}, {Stello}, {White}, {Boss}, {Devore},
  {Gould}, {Prsa}, {Agol}, {Barclay}, {Coughlin}, {Brugamyer}, {Mullally},
  {Quintana}, {Still}, {Thompson}, {Morrison}, {Twicken}, {D{\'e}sert},
  {Carter}, {Crepp}, {H{\'e}brard}, {Santerne}, {Moutou}, {Sobeck}, {Hudgins},
  {Haas}, {Robertson}, {Lillo-Box}, \& {Barrado}}]{Marcy2014}
{Marcy}, G.~W., {Isaacson}, H., {Howard}, A.~W., {et~al.} 2014, \apjs, 210, 20

\bibitem[{{Mashonkina} {et~al.}(2011){Mashonkina}, {Gehren}, {Shi}, {Korn}, \&
  {Grupp}}]{Mashonkina2011}
{Mashonkina}, L., {Gehren}, T., {Shi}, J.-R., {Korn}, A.~J., \& {Grupp}, F.
  2011, \aap, 528, A87

\bibitem[{{Mathur} {et~al.}(2011){Mathur}, {Handberg}, {Campante},
  {Garc{\'{\i}}a}, {Appourchaux}, {Bedding}, {Mosser}, {Chaplin}, {Ballot},
  {Benomar}, {Bonanno}, {Corsaro}, {Gaulme}, {Hekker}, {R{\'e}gulo},
  {Salabert}, {Verner}, {White}, {Brand{\~a}o}, {Creevey}, {Do{\v g}an},
  {Elsworth}, {Huber}, {Hale}, {Houdek}, {Karoff}, {Metcalfe},
  {Molenda-{\.Z}akowicz}, {Monteiro}, {Thompson}, {Christensen-Dalsgaard},
  {Gilliland}, {Kawaler}, {Kjeldsen}, {Quintana}, {Sanderfer}, \&
  {Seader}}]{Mathur2011}
{Mathur}, S., {Handberg}, R., {Campante}, T.~L., {et~al.} 2011, \apj, 733, 95

\bibitem[{{Molenda-{\.Z}akowicz} {et~al.}(2013){Molenda-{\.Z}akowicz}, {Sousa},
  {Frasca}, {Uytterhoeven}, {Briquet}, {Van Winckel}, {Drobek}, {Niemczura},
  {Lampens}, {Lykke}, {Bloemen}, {Gameiro}, {Jean}, {Volpi}, {Gorlova},
  {Mortier}, {Tsantaki}, \& {Raskin}}]{Molenda2013}
{Molenda-{\.Z}akowicz}, J., {Sousa}, S.~G., {Frasca}, A., {et~al.} 2013,
  \mnras, 434, 1422

\bibitem[{{Mosser} \& {Appourchaux}(2009)}]{Mosser2009}
{Mosser}, B., \& {Appourchaux}, T. 2009, \aap, 508, 877

\bibitem[{{Mosser} {et~al.}(2012){Mosser}, {Goupil}, {Belkacem}, {Michel},
  {Stello}, {Marques}, {Elsworth}, {Barban}, {Beck}, {Bedding}, {De Ridder},
  {Garc{\'{\i}}a}, {Hekker}, {Kallinger}, {Samadi}, {Stumpe}, {Barclay}, \&
  {Burke}}]{Mosser2012}
{Mosser}, B., {Goupil}, M.~J., {Belkacem}, K., {et~al.} 2012, \aap, 540, A143

\bibitem[{{Piersanti} {et~al.}(2004){Piersanti}, {Tornamb{\'e}}, \&
  {Castellani}}]{Piersanti2004}
{Piersanti}, L., {Tornamb{\'e}}, A., \& {Castellani}, V. 2004, \mnras, 353, 243

\bibitem[{{Pinsonneault} {et~al.}(2012){Pinsonneault}, {An},
  {Molenda-{\.Z}akowicz}, {Chaplin}, {Metcalfe}, \&
  {Bruntt}}]{Pinsonneault2012}
{Pinsonneault}, M.~H., {An}, D., {Molenda-{\.Z}akowicz}, J., {et~al.} 2012,
  \apjs, 199, 30

\bibitem[{{Prugniel} \& {Soubiran}(2001)}]{Prungniel2001}
{Prugniel}, P., \& {Soubiran}, C. 2001, \aap, 369, 1048

\bibitem[{{Reddy} {et~al.}(2003){Reddy}, {Tomkin}, {Lambert}, \& {Allende
  Prieto}}]{Reddy2003}
{Reddy}, B.~E., {Tomkin}, J., {Lambert}, D.~L., \& {Allende Prieto}, C. 2003,
  \mnras, 340, 304

\bibitem[{{Reimers}(1975)}]{Reimers1975}
{Reimers}, D. 1975, Memoires of the Societe Royale des Sciences de Liege, 8,
  369

\bibitem[{{Ren} {et~al.}(2016){Ren}, {Liu}, {Xiang}, {Huang}, {Hekker}, {Wang},
  {Yuan}, {Rebassa-Mansergas}, {Chen}, {Sun}, {Zhang}, {Huo}, {Zhang}, {Zhang},
  {Hou}, \& {Wang}}]{Ren2016}
{Ren}, J.-J., {Liu}, X.-W., {Xiang}, M.-S., {et~al.} 2016, Research in
  Astronomy and Astrophysics, 16, 009

\bibitem[{{S{\'a}nchez-Bl{\'a}zquez} {et~al.}(2006){S{\'a}nchez-Bl{\'a}zquez},
  {Peletier}, {Jim{\'e}nez-Vicente}, {Cardiel}, {Cenarro},
  {Falc{\'o}n-Barroso}, {Gorgas}, {Selam}, \& {Vazdekis}}]{Sanchez2006}
{S{\'a}nchez-Bl{\'a}zquez}, P., {Peletier}, R.~F., {Jim{\'e}nez-Vicente}, J.,
  {et~al.} 2006, \mnras, 371, 703

\bibitem[{{Silva Aguirre} {et~al.}(2015){Silva Aguirre}, {Davies}, {Basu},
  {Christensen-Dalsgaard}, {Creevey}, {Metcalfe}, {Bedding}, {Casagrande},
  {Handberg}, {Lund}, {Nissen}, {Chaplin}, {Huber}, {Serenelli}, {Stello}, {Van
  Eylen}, {Campante}, {Elsworth}, {Gilliland}, {Hekker}, {Karoff}, {Kawaler},
  {Kjeldsen}, \& {Lundkvist}}]{SilvaAguirre2015}
{Silva Aguirre}, V., {Davies}, G.~R., {Basu}, S., {et~al.} 2015, \mnras, 452,
  2127

\bibitem[{{Soderblom}(2010)}]{Soderblom2010}
{Soderblom}, D.~R. 2010, \araa, 48, 581

\bibitem[{{Steinmetz} {et~al.}(2006){Steinmetz}, {Zwitter}, {Siebert},
  {Watson}, {Freeman}, {Munari}, {Campbell}, {Williams}, {Seabroke}, {Wyse},
  {Parker}, {Bienaym{\'e}}, {Roeser}, {Gibson}, {Gilmore}, {Grebel}, {Helmi},
  {Navarro}, {Burton}, {Cass}, {Dawe}, {Fiegert}, {Hartley}, {Russell},
  {Saunders}, {Enke}, {Bailin}, {Binney}, {Bland-Hawthorn}, {Boeche}, {Dehnen},
  {Eisenstein}, {Evans}, {Fiorucci}, {Fulbright}, {Gerhard}, {Jauregi}, {Kelz},
  {Mijovi{\'c}}, {Minchev}, {Parmentier}, {Pe{\~n}arrubia}, {Quillen}, {Read},
  {Ruchti}, {Scholz}, {Siviero}, {Smith}, {Sordo}, {Veltz}, {Vidrih}, {von
  Berlepsch}, {Boyle}, \& {Schilbach}}]{Steinmetz2006}
{Steinmetz}, M., {Zwitter}, T., {Siebert}, A., {et~al.} 2006, \aj, 132, 1645

\bibitem[{{Stello} {et~al.}(2013){Stello}, {Huber}, {Bedding}, {Benomar},
  {Bildsten}, {Elsworth}, {Gilliland}, {Mosser}, {Paxton}, \&
  {White}}]{Stello2013}
{Stello}, D., {Huber}, D., {Bedding}, T.~R., {et~al.} 2013, \apjl, 765, L41

\bibitem[{{Takeda} \& {Tajitsu}(2015)}]{Takeda2015}
{Takeda}, Y., \& {Tajitsu}, A. 2015, \mnras, 450, 397

\bibitem[{{Thygesen} {et~al.}(2012){Thygesen}, {Frandsen}, {Bruntt},
  {Kallinger}, {Andersen}, {Elsworth}, {Hekker}, {Karoff}, {Stello},
  {Brogaard}, {Burke}, {Caldwell}, \& {Christiansen}}]{Thygesen2012}
{Thygesen}, A.~O., {Frandsen}, S., {Bruntt}, H., {et~al.} 2012, \aap, 543, A160

\bibitem[{{Ulrich}(1986)}]{Ulrich1986}
{Ulrich}, R.~K. 1986, \apjl, 306, L37

\bibitem[{{van Leeuwen}(2007)}]{vanLeeuwen2007}
{van Leeuwen}, F. 2007, \aap, 474, 653

\bibitem[{{Verner} {et~al.}(2011){Verner}, {Chaplin}, {Basu}, {Brown},
  {Hekker}, {Huber}, {Karoff}, {Mathur}, {Metcalfe}, {Mosser}, {Quirion},
  {Appourchaux}, {Bedding}, {Bruntt}, {Campante}, {Elsworth}, {Garc{\'{\i}}a},
  {Handberg}, {R{\'e}gulo}, {Roxburgh}, {Stello}, {Christensen-Dalsgaard},
  {Gilliland}, {Kawaler}, {Kjeldsen}, {Allen}, {Clarke}, \&
  {Girouard}}]{Verner2011}
{Verner}, G.~A., {Chaplin}, W.~J., {Basu}, S., {et~al.} 2011, \apjl, 738, L28

\bibitem[{{Vogt} {et~al.}(1994){Vogt}, {Allen}, {Bigelow}, {Bresee}, {Brown},
  {Cantrall}, {Conrad}, {Couture}, {Delaney}, {Epps}, {Hilyard}, {Hilyard},
  {Horn}, {Jern}, {Kanto}, {Keane}, {Kibrick}, {Lewis}, {Osborne},
  {Pardeilhan}, {Pfister}, {Ricketts}, {Robinson}, {Stover}, {Tucker}, {Ward},
  \& {Wei}}]{Vogt1994}
{Vogt}, S.~S., {Allen}, S.~L., {Bigelow}, B.~C., {et~al.} 1994, in Society of
  Photo-Optical Instrumentation Engineers (SPIE) Conference Series, Vol. 2198,
  Instrumentation in Astronomy VIII, ed. D.~L. {Crawford} \& E.~R. {Craine},
  362

\bibitem[{{Wan} {et~al.}(2015){Wan}, {Liu}, {Deng}, {Cui}, {Zhang}, {Hou},
  {Yang}, \& {Wu}}]{Wan2015}
{Wan}, J.-C., {Liu}, C., {Deng}, L.-C., {et~al.} 2015, Research in Astronomy
  and Astrophysics, 15, 1166

\bibitem[{{Wang} {et~al.}(2011){Wang}, {Liu}, {Zhao}, \& {Sato}}]{Wang2011}
{Wang}, L., {Liu}, Y., {Zhao}, G., \& {Sato}, B. 2011, \pasj, 63, 1035

\bibitem[{{Wu} {et~al.}(2014){Wu}, {Du}, {Luo}, {Zhao}, \& {Yuan}}]{Wu2014}
{Wu}, Y., {Du}, B., {Luo}, A., {Zhao}, Y., \& {Yuan}, H. 2014, in IAU
  Symposium, Vol. 306, Statistical Challenges in 21st Century Cosmology, ed.
  A.~{Heavens}, J.-L. {Starck}, \& A.~{Krone-Martins}, 340--342

\bibitem[{{Wu} {et~al.}(2011){Wu}, {Luo}, {Li}, {Shi}, {Prugniel}, {Liang},
  {Zhao}, {Zhang}, {Bai}, {Wei}, {Dong}, {Zhang}, \& {Chen}}]{Wu2011}
{Wu}, Y., {Luo}, A.-L., {Li}, H.-N., {et~al.} 2011, Research in Astronomy and
  Astrophysics, 11, 924

\bibitem[{{Xiang} {et~al.}(2015){Xiang}, {Liu}, {Yuan}, {Huang}, {Huo},
  {Zhang}, {Chen}, {Zhang}, {Sun}, {Wang}, {Zhao}, {Shi}, {Luo}, {Li}, {Wu},
  {Bai}, {Zhang}, {Hou}, {Yuan}, {Li}, \& {Wei}}]{Xiang2015}
{Xiang}, M.~S., {Liu}, X.~W., {Yuan}, H.~B., {et~al.} 2015, \mnras, 448, 822

\bibitem[{{Xue} {et~al.}(2014){Xue}, {Ma}, {Rix}, {Morrison}, {Harding},
  {Beers}, {Ivans}, {Jacobson}, {Johnson}, {Lee}, {Lucatello}, {Rockosi},
  {Sobeck}, {Yanny}, {Zhao}, \& {Allende Prieto}}]{Xue2014}
{Xue}, X.-X., {Ma}, Z., {Rix}, H.-W., {et~al.} 2014, \apj, 784, 170

\bibitem[{{Yanny} {et~al.}(2009){Yanny}, {Rockosi}, {Newberg}, {Knapp},
  {Adelman-McCarthy}, {Alcorn}, {Allam}, {Allende Prieto}, {An}, {Anderson},
  {Anderson}, {Bailer-Jones}, {Bastian}, {Beers}, {Bell}, {Belokurov},
  {Bizyaev}, {Blythe}, {Bochanski}, {Boroski}, {Brinchmann}, {Brinkmann},
  {Brewington}, {Carey}, {Cudworth}, {Evans}, {Evans}, {Gates}, {G{\"a}nsicke},
  {Gillespie}, {Gilmore}, {Nebot Gomez-Moran}, {Grebel}, {Greenwell}, {Gunn},
  {Jordan}, {Jordan}, {Harding}, {Harris}, {Hendry}, {Holder}, {Ivans},
  {Ivezi{\v c}}, {Jester}, {Johnson}, {Kent}, {Kleinman}, {Kniazev},
  {Krzesinski}, {Kron}, {Kuropatkin}, {Lebedeva}, {Lee}, {French Leger},
  {L{\'e}pine}, {Levine}, {Lin}, {Long}, {Loomis}, {Lupton}, {Malanushenko},
  {Malanushenko}, {Margon}, {Martinez-Delgado}, {McGehee}, {Monet}, {Morrison},
  {Munn}, {Neilsen}, {Nitta}, {Norris}, {Oravetz}, {Owen}, {Padmanabhan},
  {Pan}, {Peterson}, {Pier}, {Platson}, {Re Fiorentin}, {Richards}, {Rix},
  {Schlegel}, {Schneider}, {Schreiber}, {Schwope}, {Sibley}, {Simmons},
  {Snedden}, {Allyn Smith}, {Stark}, {Stauffer}, {Steinmetz}, {Stoughton},
  {SubbaRao}, {Szalay}, {Szkody}, {Thakar}, {Sivarani}, {Tucker}, {Uomoto},
  {Vanden Berk}, {Vidrih}, {Wadadekar}, {Watters}, {Wilhelm}, {Wyse}, {Yarger},
  \& {Zucker}}]{Yanny2009}
{Yanny}, B., {Rockosi}, C., {Newberg}, H.~J., {et~al.} 2009, \aj, 137, 4377

\bibitem[{{Zwitter} {et~al.}(2004){Zwitter}, {Castelli}, \&
  {Munari}}]{Zwitter2004}
{Zwitter}, T., {Castelli}, F., \& {Munari}, U. 2004, \aap, 417, 1055

\bibitem[{{Zwitter} {et~al.}(2008){Zwitter}, {Siebert}, {Munari}, {Freeman},
  {Siviero}, {Watson}, {Fulbright}, {Wyse}, {Campbell}, {Seabroke}, {Williams},
  {Steinmetz}, {Bienaym{\'e}}, {Gilmore}, {Grebel}, {Helmi}, {Navarro},
  {Anguiano}, {Boeche}, {Burton}, {Cass}, {Dawe}, {Fiegert}, {Hartley},
  {Russell}, {Veltz}, {Bailin}, {Binney}, {Bland-Hawthorn}, {Brown}, {Dehnen},
  {Evans}, {Re Fiorentin}, {Fiorucci}, {Gerhard}, {Gibson}, {Kelz}, {Kujken},
  {Matijevi{\v c}}, {Minchev}, {Parker}, {Pe{\~n}arrubia}, {Quillen}, {Read},
  {Reid}, {Roeser}, {Ruchti}, {Scholz}, {Smith}, {Sordo}, {Tolstoi},
  {Tomasella}, {Vidrih}, \& {Wylie-de Boer}}]{Zwitter2008}
{Zwitter}, T., {Siebert}, A., {Munari}, U., {et~al.} 2008, \aj, 136, 421

\end{thebibliography}

\end{document}